\documentclass[aps,prl,twocolumn,superscriptaddress,showpacs]{revtex4-1}

\usepackage{graphicx}

\begin{document}
\title{Chirality-induced asymmetric magnetic nucleation in Pt/Co/AlOx ultrathin microstructures}

\author{S.~Pizzini } \email{stefania.pizzini@neel.cnrs.fr}
\affiliation{CNRS, Institut N\'{e}el, 38042 Grenoble, France}
\affiliation{Univ.~Grenoble Alpes, Institut N\'{e}el, 38042 Grenoble, France}
\author{J.~Vogel}
\affiliation{CNRS, Institut N\'{e}el, 38042 Grenoble, France}
\affiliation{Univ.~Grenoble Alpes, Institut N\'{e}el, 38042 Grenoble, France}
\author{S.~Rohart}
\affiliation{Laboratoire de Physique des Solides, Univ. Paris-Sud, CNRS UMR 8502, 91405 Orsay, France}
\author{L.D.~Buda-Prejbeanu}
\affiliation{SPINTEC, UMR 8191, CEA/CNRS/UJF/Grenoble-INP, INAC, 38054
Grenoble Cedex, France}
\author{E.~Ju\'{e}}
\affiliation{SPINTEC, UMR 8191, CEA/CNRS/UJF/Grenoble-INP, INAC, 38054
Grenoble Cedex, France}
\author{O.~Boulle}
\affiliation{SPINTEC, UMR 8191, CEA/CNRS/UJF/Grenoble-INP, INAC, 38054
Grenoble Cedex, France}
\author{I.M.~Miron}
\affiliation{SPINTEC, UMR 8191, CEA/CNRS/UJF/Grenoble-INP, INAC, 38054
Grenoble Cedex, France}
\author{C.K.~Safeer}
\affiliation{SPINTEC, UMR 8191, CEA/CNRS/UJF/Grenoble-INP, INAC, 38054
Grenoble Cedex, France}
\author{S.~Auffret}
\affiliation{SPINTEC, UMR 8191, CEA/CNRS/UJF/Grenoble-INP, INAC, 38054
Grenoble Cedex, France}
\author{G.~Gaudin}
\affiliation{SPINTEC, UMR 8191, CEA/CNRS/UJF/Grenoble-INP, INAC, 38054
Grenoble Cedex, France}
\author{A.~Thiaville}
\affiliation{Laboratoire de Physique des Solides, Univ. Paris-Sud, CNRS UMR 8502, 91405 Orsay, France}

\pacs{75.70.Ak, 75.60.Jk}

\begin{abstract}
The nucleation of reversed magnetic domains in Pt/Co/AlO$_{x}$ microstructures with perpendicular
anisotropy was studied experimentally in the presence of an in-plane magnetic field.
For large enough in-plane field, nucleation was observed preferentially  at an edge of the sample normal to this field.
The  position at which nucleation takes place was observed to depend in a chiral way on the initial
 magnetization and applied field directions.
A quantitative explanation of these results is proposed, based on the existence of a sizable Dzyaloshinskii-Moriya
interaction in this sample.
Another consequence of this interaction is that the energy of domain walls can become negative for in-plane fields smaller
than the effective anisotropy field.
\end{abstract}

\maketitle

Chirality is a fascinating property of nature \cite{Mislow1999}.
It was discovered in 1848 by Pasteur, by correlating the optical activity
of molecules in solution to the hemihedral shape of the crystals that they form \cite{note-chiralite}.
More generally, chiral textures often appear in physics as a result of symmetry breaking, either
spontaneously for example in macroscopic quantum systems \cite{Volovik2012}, or when
stabilized by a chiral interaction as in liquid crystals \cite{deGennes1993} or magnetism \cite{Dzyaloshinskii1957}.
In the latter case, the transcription of the chirality from the atomic scale to the macroscopic scale of
textures may be impeded by the existence of an anisotropy.
This is exemplified in liquid crystals, where under a DC magnetic or electric field that induces anisotropy, the
cholesteric-nematic transition takes place \cite{deGennes1993}.
Thus, the detection and quantification of a chiral interaction when it is too weak to give rise to a global chiral
texture is difficult.

Magnetism is another prominent field where chiral textures are considered.
A chiral magnetic interaction indeed exists, namely an anti-symmetric exchange called Dzyaloshinskii-Moriya
interaction (DMI) \cite{Dzyaloshinskii1957,Moriya1960}, that is allowed when central symmetry is broken.
In many compounds having this property, especially the cubic B20 structures with depressed magnetic anisotropy,
chiral textures like homochiral spin spirals or 2D skyrmion lattices have been
observed, both in reciprocal \cite{Muhlbauer2009} and in real space \cite{Yu2010}.
Another class of chiral magnetic systems has recently appeared, namely the few atomic layer thick samples
grown on an underlayer with large spin-orbit interaction, showing structural inversion asymmetry and perpendicular
magnetic anisotropy (PMA) \cite{Bode2007,Meckler2009,Miron2010,Miron2011}.
In these systems, PMA is very strong so that chiral spin spirals are not
stable, and as a result DMI has remained unnoticed for about 20 years.
However, at  magnetic edges like a domain wall (DW) separating two uniformly magnetized domains or at physical
edges in a microstructure, the competition of chiral interaction with anisotropy is modified.
Indeed, the peculiarities of field and current-induced dynamics of domain walls in such samples
\cite{Emori2013,Ryu2013,Je2013} have been found to be consistent with a chiral texture localized on the DW, deriving
from the presence of interface-induced DMI \cite{Thiaville2012}.
These local chiral magnetization textures, that appear as N\'{e}el walls of a fixed chirality,
have also been observed recently by low-energy electron microscopy \cite{Chen2013,Chen2013b}, on samples
with wide domain walls.

In this Letter we show that chiral interactions can also be detected at the edges of a microstructure:
in the presence of an additional in-plane field, nucleation of reversed domains takes place
preferentially at one edge of the sample, oriented perpendicular to this field.
The side at which nucleation takes place depends on the direction of both the additional field and the initial
magnetization.
This asymmetry is thus chiral, and we show that DMI can explain this chirality  as well as the values of the nucleation field.

The experiments were carried out on Pt(3~nm)/Co(0.6~nm)/AlO$_{x}$(2~nm) layers patterned by electron beam lithography
into two large injection pads connected by micrometric strips.
The strips were used for field and current-driven domain wall dynamics (not shown here) while the nucleation
experiments were carried out on the pads.
The films were deposited on a Si/SiO$_{2}$ substrate by magnetron sputtering.
Samples were oxidized \textit{in situ} by oxygen plasma in order to induce PMA \cite{Manchon2008b}.
Magnetization reversal was studied using  magneto-optical Kerr microscopy.
In each experiment, magnetization was first saturated with an out-of-plane  magnetic field ($H_{z}$).
Nucleation of reversed domains was then induced by an opposite $H_{z}$ field pulse, under a DC
in-plane field $H_x$.
Several samples with similar composition and varying magnetic anisotropy were measured.

\begin{figure}[ht!]
\includegraphics[width=8.5cm]{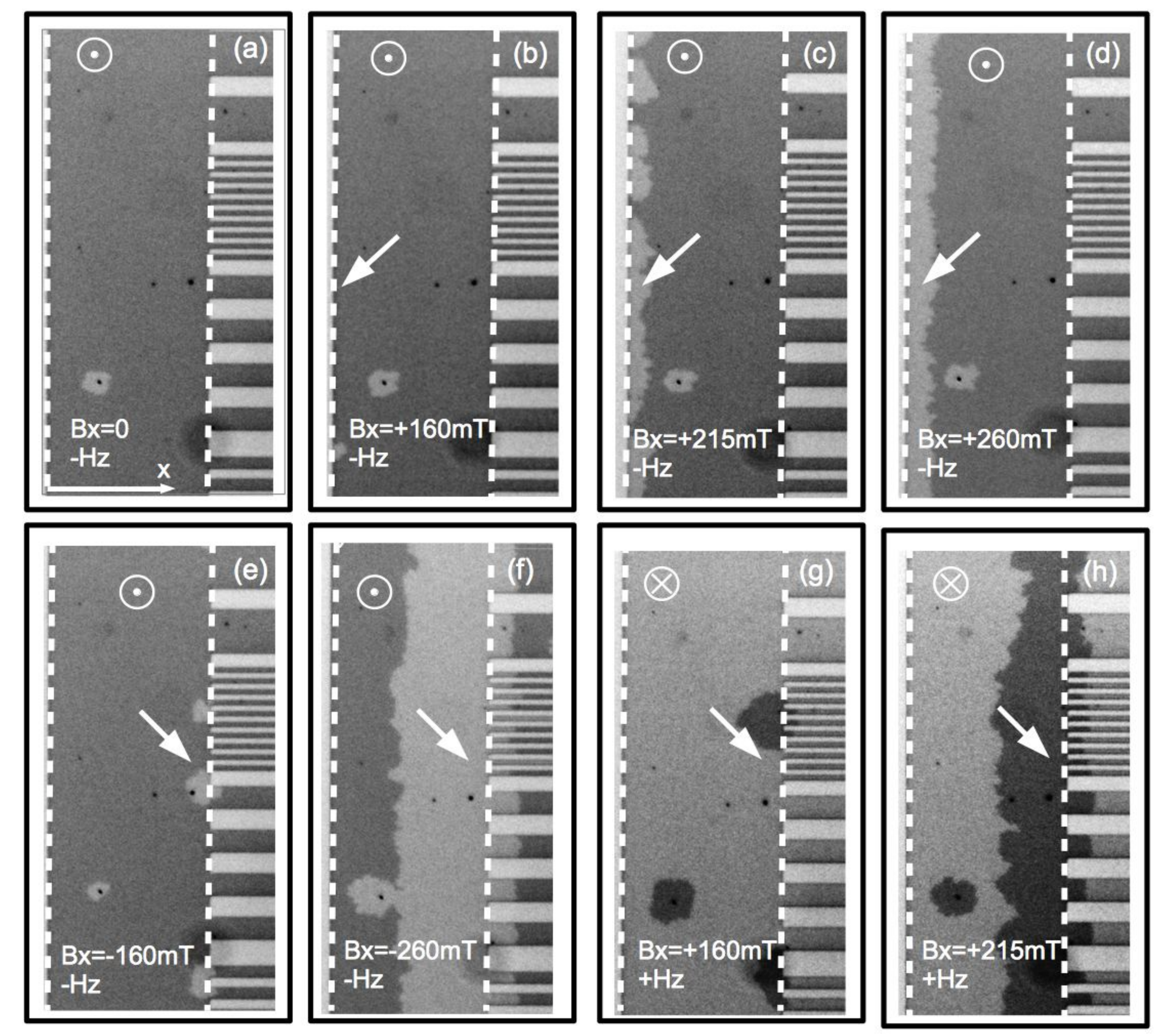}
\caption{\label{fig:nucl-fig1a} Kerr images showing the chiral nucleation of domains at one edge of
the pad of the Pt/Co/AlO$_{x}$ microstructure, by application of an out-of-plane field pulse.
(a)-(d): magnetization is initially saturated $\uparrow$ and $B_{x}$=0, +160, +215 and +260~mT;
(e-f) magnetization is initially saturated $\uparrow$ and $B_{x}$ =-160~mT and -260~mT;
(g)-(h) magnetization is initially saturated $\downarrow$ and $B_{x}$ is
+160~mT and +260~mT. The width of the pad is 70~$\mu$m.
The dotted lines highlight the left and right edges of the pad and the arrows show the side of the
sample where nucleation takes place.}
\end{figure}

Figure~\ref{fig:nucl-fig1a} illustrates an example of the occurrence of chiral nucleation and the symmetry of this effect.
When an $H_{z}$ field pulse (amplitude $18-20$~mT and length $50-100$~ms)
is applied antiparallel to the initial  magnetization direction,
magnetization reversal is initiated by the nucleation of a reversed domain at a particular spot
of the sample, away from the edges, corresponding a local defect (Fig.~\ref{fig:nucl-fig1a}(a)).
When a sufficiently strong in-plane field $H_{x}$ is applied at the same time as $H_{z}$, new
nucleation centers appear at one edge of the pad.
Starting from positive ($\uparrow$) saturation
(corresponding to dark contrast in the Kerr images), a positive $H_{x}$ field (along the positive $x$ axis)
promotes nucleation of reversed $\downarrow$ domains at the left edge of the sample
(Fig.~\ref{fig:nucl-fig1a}(c-e)).
As the amplitude of $H_{x}$ increases, the nucleation probability increases but no nucleation
appears at the right edge of the sample, up to $\mu_{0}H_x = 260$~mT. If either the initial magnetization direction or the $H_{x}$ field direction is reversed, the nucleation
takes place at the opposite edge.
Figures \ref{fig:nucl-fig1a}(e-f) show indeed that nucleation takes place at the right edge when a
negative $H_{x}$ field is applied starting from the same $\uparrow$ saturation.
Similarly, nucleation takes place on the right edge when $H_{x}$ is kept positive but the initial
magnetization is reversed ($\downarrow$) (Fig.~\ref{fig:nucl-fig1a}(g-h)).
This shows that the observed asymmetry is indeed chiral.

\begin{figure}[ht!]
\includegraphics[width=8.5cm]{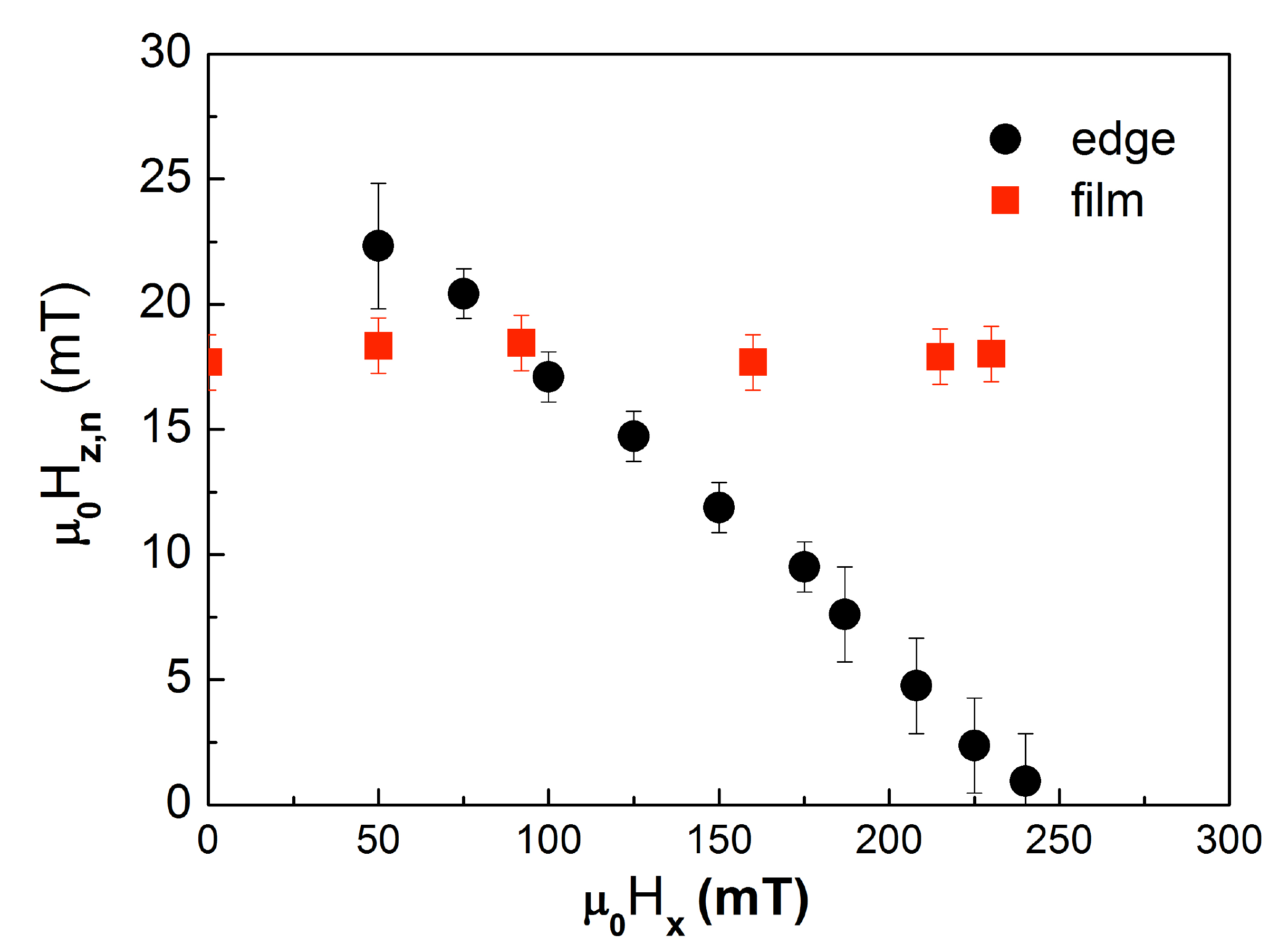}
\caption{\label{fig:Figure2} (color online). Nucleation fields measured as a function of  in-plane field $H_{x}$
for the reversed domain in the middle (squares) and at the left edge of the sample (circles). Note the difference of scales between the two axes.}
\end{figure}

The nucleation field $H_{z,n}$ was measured  as a function of $H_{x}$ for a sample having an anisotropy field of 700~mT (slightly weaker than that of the sample shown in Fig.~\ref{fig:nucl-fig1a}) both for a defect within the
film and at the sample edges. The length of the $H_{z}$ pulse was fixed at 20~ms.
For a defect within the film, $H_{z,n}$ was defined as the field for which the domain appears with 100\% probability.
For the edges, $H_{z,n}$ was defined as the field for which 10-15 domains systematically nucleate \cite{SI-NucleDM}.
The main result of these measurements, shown in Fig.~\ref{fig:Figure2}, is that while the nucleation field of a domain
within the film is almost $H_{x}$-field independent, $H_{z,n}$
strongly decreases with the in-plane field for the domains nucleating at the sample edges.

The chiral behavior of magnetization reversal cannot be explained by
simply invoking a local reduction of anisotropy along the sample edges,
which would keep the symmetry between opposite edges.
In order to explain the observed chiral nucleation, a phenomenon which breaks the
symmetry of the system when an $H_{x}$ field is applied has to be invoked.
A possible origin of this phenomenon is the presence in
non-centrosymmetric Pt/Co/AlO$_{x}$ stacks of a non-vanishing DM
interaction, which has already been invoked to explain the stabilization of chiral N\'{e}el walls, called Dzyaloshinskii domain walls (DDW) \cite{Thiaville2012}.
We thus quantitatively investigate this hypothesis using two models.

\par
\begin{figure}[ht!]
\includegraphics[width=8.5cm]{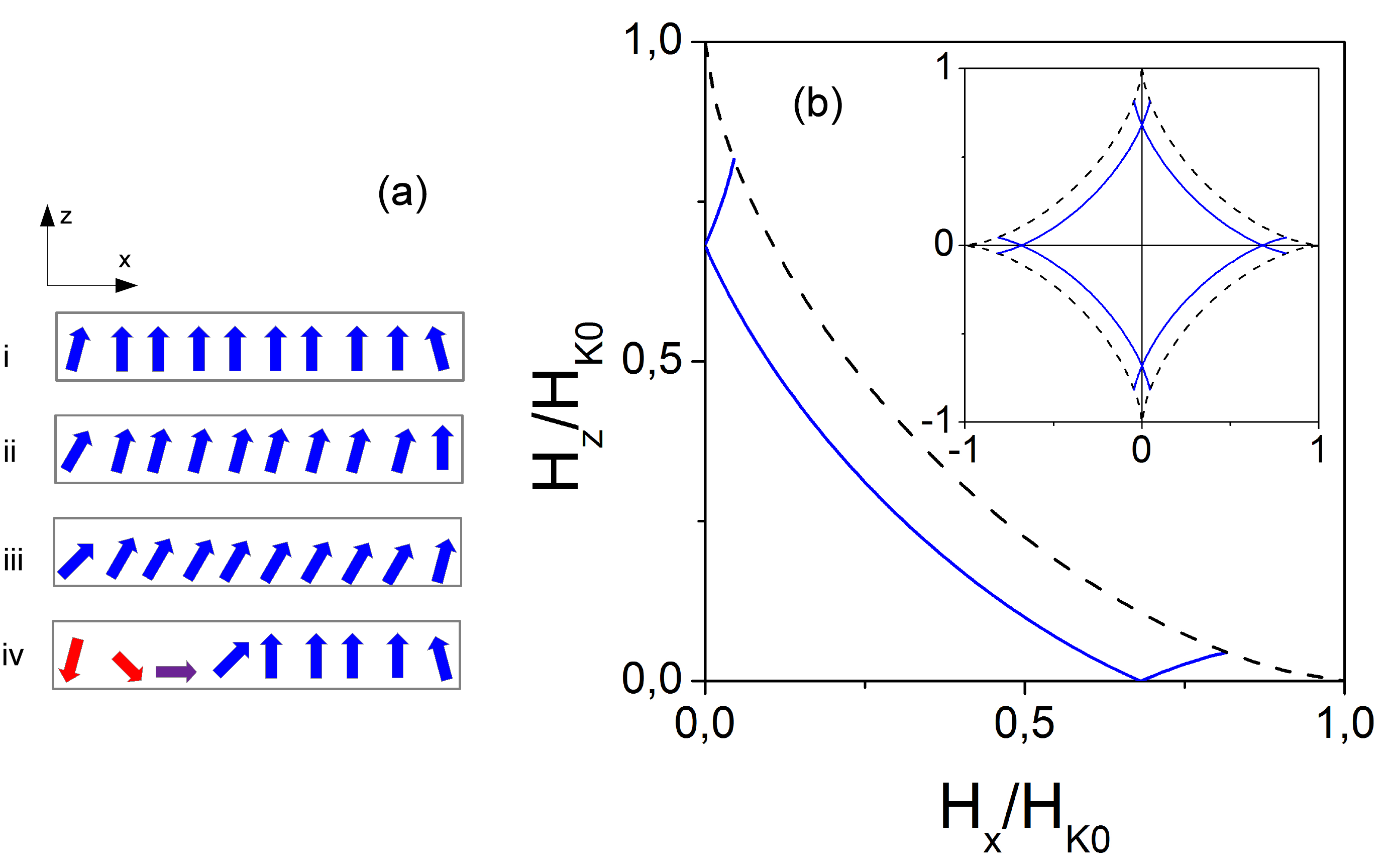}
\caption{\label{fig:nucl_fig3} (color online). a) Sketch of the
micromagnetic configuration within a microstructure with DMI in zero
applied field (i), under an $x$ field (ii), under an additional negative
$z$ field (iii), and after reversal, with a domain wall of magnetization
parallel to the $x$ field (iv).
b) results of a 1D calculation showing the
reversal field for $D/D_{c0}= 0$ (dashes) and 0.5 (lines).
For $D\neq0$ an easy and a hard branch develop, corresponding to the
reversal at the two edges of the microstructure.
Inset: complete astroids.}
\end{figure}

{\emph{Zero temperature model.}}
In real samples (i.e. including defects), magnetization reversal is
controlled by few defects acting as nucleation centers \cite{Aharoni1962}.
Chiral nucleation requires defects with a chiral micromagnetic structure around them.
DMI provides such a state at sample
edges, inducing locally a tilt of the magnetization \cite{Rohart2013}.
When an in-plane field normal to an edge is applied, the tilt angle depends on
its orientation (parallel or anti-parallel to the field) and preferential
nucleation at one edge can be expected.
Using the same 1D model in the $x$ direction normal to the edge as in \cite{Rohart2013},
the edge tilt angle $\theta$ is given by
\begin{equation} \label{eq:theta-boundary}
m_{x}=\sin \theta=\pm \Delta \frac{D}{2A} + \frac{H_{x}}{H_{K0}},
\end{equation}
where $D$ is the DMI constant, $A$ the exchange constant,
$H_{K0}=2K_0/(\mu_{0}M_{s})$ the anisotropy field ($K_0$ is the
effective anisotropy constant), $\Delta=\sqrt{A/K_0}$ the nominal domain
wall width and the $\pm$ sign refers to the two edges of the sample
along $x$.
Figure \ref{fig:nucl_fig3}(a) sketches the effect of $H_{x}$ and $H_{z}$
on the micromagnetic configurations.

In the absence of thermal fluctuations, the solution for the onset of
magnetization instability at the edge can be mapped \cite{SI-NucleDM} to a solution of the
Stoner-Wohlfarth model \cite{Stoner1948}.
Figure~\ref{fig:nucl_fig3}(b) shows the reversal field $H_{z}$ versus $H_{x}$
(normalized to $H_{K0}$), calculated for  different $D/D_{c0}$
(with $D_{c0} = 4 \sqrt{A K_0}/\pi \equiv \sigma_{00}/\pi$ giving the onset of the spontaneous formation
of magnetization cycloids ).
For $D=0$,  the standard Stoner-Wohlfarth astroid is obtained and no difference occurs between
the sample edge and the center.
For a finite $D$, the edge reversal field with $H_{x}=0$ decreases by a factor
 $[1-(2/\pi)D/D_{c0}]$.
As $H_{x}$ is applied, the astroid splits into two branches revealing the
difference between the two sample sides: on the side where $m_{x}$ is
initially parallel (resp. anti-parallel) to $H_{x}$ the tilt is larger
(resp. smaller) and the reversal field decreases (resp. increases) with
$H_x$.

This model is in qualitative agreement with the experimental results:
\textit{i)} it explains why in the presence of DMI magnetic nucleation
is observed only at one sample edge, \textit{ii)} it explains
qualitatively the decrease of nucleation field as the $H_{x}$ field
amplitude is increased. However, the calculated values of the nucleation
field are about one order of magnitude larger than the experimental ones.

\begin{figure}[ht!]
\includegraphics[width=8.5cm]{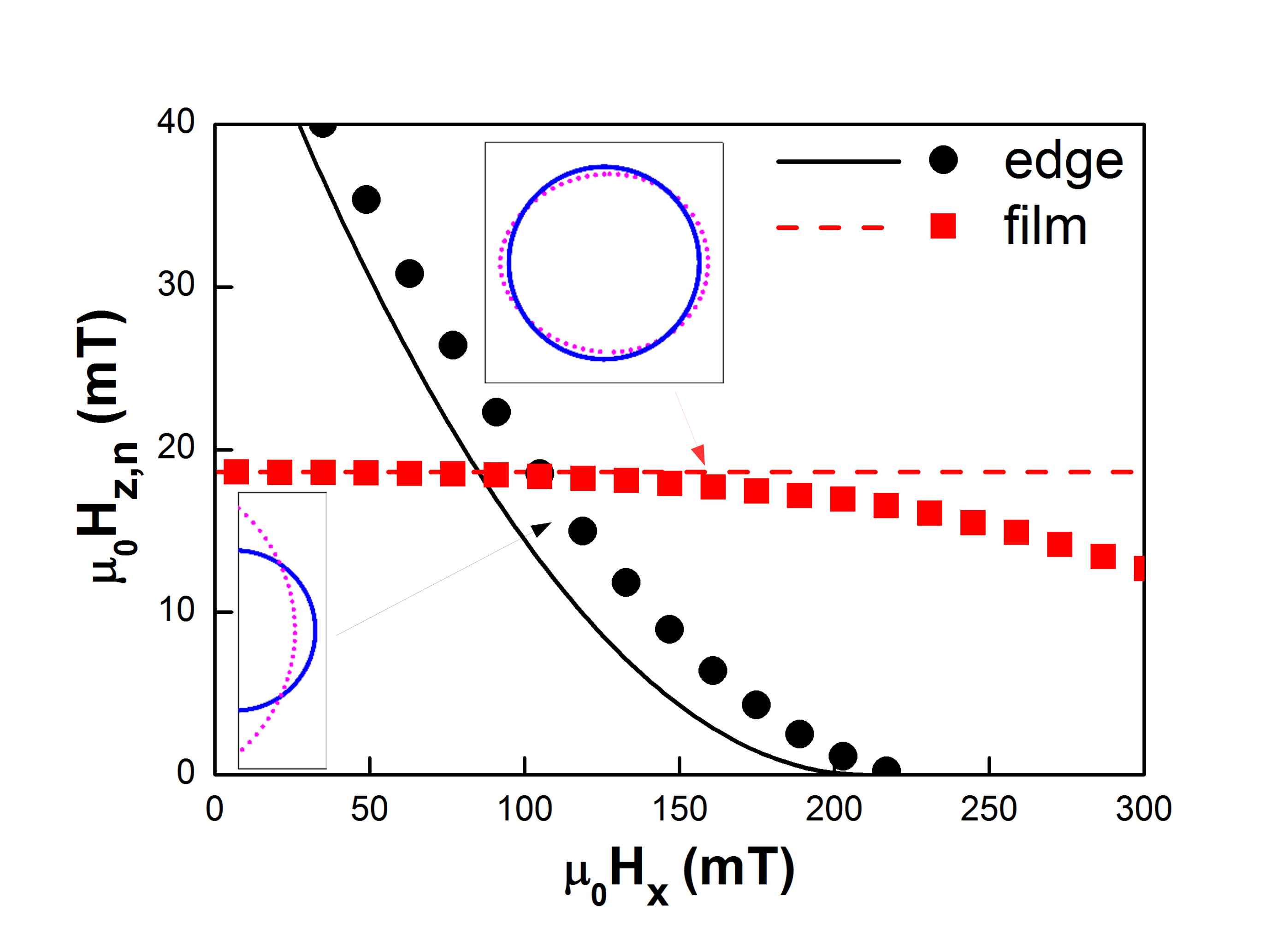}
\caption{\label{fig:nucl-fig4} (color online). Nucleation field \textit{vs} $H_{x}$ for a domain at the edge
and in the film calculated using the rigid droplet model (full and dotted line) and the relaxed model (dots and square symbols).
Insets: calculated droplet shapes for $H_x/H_{K0}=0$ (lines) and 0.21 (dots).}
\end{figure}

\par{\emph{Finite temperature model.}}
In a macroscopic sample, magnetization reversal occurs \textit{via} the creation of reversed domains followed by
the propagation of domain walls.
This is described by the so-called `droplet model' \cite{Barbara1976,Barbara1994} well-known for first order
phase transitions.
Let us first consider the creation of a cylindrical domain of radius $R$ inside the film.
The free energy of this droplet is:
\begin{equation}\label{eq:droplet-film}
E/t=2\pi R \sigma_{0} - 2 \mu_{0}M_{s}H_{z}\pi R^{2},
\end{equation}
where $t$ is the film thickness, $\sigma_{0}=\sigma_{00} (1 - D/D_{c0})$ the domain wall energy density in the
presence of DMI \cite{Thiaville2012}, and $H_{z}$ the applied magnetic field.
The critical droplet radius is $R_{c}=\sigma_{0}/(2\mu_{0}M_{s}H_{z})$.
Below $R_{c}$ the droplet collapses, whereas above $R_{c}$ the domain increases its size by DW propagation.
This gives rise to an energy barrier for the nucleation of the droplet:
\begin{equation}\label{eq:droplet-barrier}
E_{B}=\frac{\pi{\sigma_{0}}^2 t}{2\mu_{0}M_{s}H_{z}}.
\end{equation}
In an Arrhenius model with attempt time $\tau_0$, the nucleation field for a waiting time
$\tau=\tau_{0}e^{p}$ reads then:
\begin{equation}\label{droplet-Hn}
H_{n,film}=\frac{\pi{\sigma_{0}}^2 t}{2\mu_{0}M_{s}pk_{B}T}
\end{equation}
Under the assumption that the magnetic droplet structure is completely rigid (\textit{i.e} no magnetization rotation in the
domains nor in the domain wall, no droplet shape optimization), the application of an in-plane field $H_{x}$ does not modify the
energy of the droplet: the Zeeman energy gained within the  half-droplet having a DW magnetization component
parallel to $H_{x}$ is compensated by the loss of energy within the half-droplet with opposite magnetization.
This agrees with the results of Fig.~\ref{fig:Figure2} which show that the nucleation field for the reversed
domain away from the edges does not vary with $H_{x}$.

On the other hand, the energy of a  half-droplet nucleating at one edge of the sample is modified by the
in-plane field.
By again assuming a rigid droplet structure, its energy reads:
\begin{equation}\label{eq:droplet-edge-1}
E/t=\pi R (\sigma_{0} \mp 2\Delta \mu_{0}M_{s}H_{x})-\mu_{0}M_{s}H_{z}\pi R^{2}
\end{equation}
where the Zeeman energy associated to the in-plane field $H_{x}$ within the DW volume has been included in
the DW energy.
The $\mp$ sign refers to the gain/loss of Zeeman energy for a domain wall having its magnetization
parallel/anti-parallel to the applied in-plane field, respectively, i.e. to the two sample edges.
In analogy with the nucleation within  the film, the nucleation field at the edges is then:
\begin{equation}\label{eq:droplet-edge-Hn}
H_{n, edge}=\frac{\pi(\sigma_{0}\mp 2\Delta \mu_{0}M_{s}H_{x})^2 t}{4\mu_{0}M_{s}pk_{B}T}
\end{equation}
This shows that the presence of DMI gives rise to different nucleation fields for DW having
magnetization parallel or anti-parallel to $H_{x}$.
In a sample with DMI, the DDW created at the two edges starting from saturation have opposite magnetization:
this explains the nucleation of reversed domains only at one side of the sample and the symmetry of the effect
when in-plane field direction and magnetization saturation are reversed.
Note that experimentally, nucleation at the hard side of the sample is never observed at larger $H_{z}$
fields, as the magnetization is always reversed by propagation of the domain walls formed at the easy side for smaller
nucleation fields.

Figure \ref{fig:nucl-fig4} (lines) shows the variation of the nucleation field as a function of $H_{x}$,
calculated for a droplet within the film and a half-droplet having its magnetization parallel to the
applied field.
The used magnetic parameters for Pt/Co/AlO$_{x}$ are $A=16$~pJ/m, M$_{s}$=1.1~MA/m, $\mu_0 H_{K0}=700$~mT,
and $D=2.2$~mJ/m$^{2}$ ($D / D_{c0}=0.7$).

It can be seen that the droplet model including the presence of large DMI provides an excellent
understanding of the measurements: not only the different variation of $H_{z,n}$ vs. $H_{x}$ within the film and
at the easy edge is accounted for, but also the order of magnitude of the reversal fields at the edge is in quantitative agreement with the experiments.
In the film, in order to account for the local reduction of the anisotropy field at the defect and therefore reproduce the experimental values of the nucleation field,
the energy of the domain wall was reduced by a factor $\epsilon \approx 0.4$
as done previously \cite{Moritz2005}.
Note that the difference between the theoretical and the experimental curve while approaching the
\textit{x}-axis can be explained by the non-perfect compensation of the tilt of the $H_{x}$ field
with respect to the sample surface.

Lifting the first two restrictions of the rigid model can be performed semi-analytically, using the `small circle' model
(the wall magnetization distribution is assumed to belong to a plane, that cuts the order parameter sphere
along a small circle).
This provides accurate DW energies, as was shown long ago \cite{Hubert1974} and checked again here.
Once the orientation-dependent DW energy is known, the optimal droplet shape is obtained using
the Wulff construction \cite{Desjonqueres1996}.
In the case of the half droplet, in full analogy with the calculation of the contact angle of a liquid
droplet on a surface, the difference of edge energies for up and down domain magnetization (that can
be analytically calculated with the same model as used in the Zero temperature section) was introduced in
the Wulff construction.
For each value of $H_x$, the droplet shapes were first computed (see insets in Fig.~\ref{fig:nucl-fig4}).
Inside the film, an asymmetric elongation along the in-plane field is seen.
At the edge, a significant elongation perpendicular to the field takes place, as the DW oriented perpendicular to the field
has a much reduced energy and can expand its length.
With the shape fixed, the determination of the critical droplet size was then performed \cite{SI-NucleDM}.
The numerical results of this semi-analytical model, shown in Fig.~\ref{fig:nucl-fig4} (symbols) for the
case $D/D_{c0}=0.7$, are very close to experiments.
The new feature is the decrease of the nucleation field at the defect in the film, as the in-plane field increases.
Comparing calculations and experimental data showed that, within this model, $D \geq 0.7 D_{c0}$
is required to get similar evolutions with in-plane field magnitude  \cite{SI-NucleDM}.

The decrease to zero of the nucleation field, seen both experimentally and in the model, is due
to the decrease to zero of the DDW energy under a sufficiently large in-plane field.
In the case of an in-plane field normal to the DDW,
the DW profile can be analytically calculated for an arbitrary value of the in-plane field.
For DMI-induced DW moment in the same direction as the field, the DW energy reads
(with $h= H_x / H_{K0}$ \cite{noteHubert}):
\begin{equation}\label{full-DW-profile}
\sigma=\sigma_{00}[\sqrt{1-h^{2}}-(h+\frac{2}{\pi}\frac{D}{D_{c0}}) \arccos h].
\end{equation}
This falls to zero at $h \approx 1 - D/D_{c0}$.
For other in-plane angles between field and DW normal, the zero crossing takes place at larger fields,
reaching $H_{K0}$ when the field is along the DW.
This appears to be a unique feature of the Dzyaloshinskii DW.

In conclusion, the nucleation of reversed domains in Pt/Co/AlO$_{x}$ microstructures was observed to be chiral, and could be
explained by the presence of a strong DM interaction, already identified as being responsible for the chiral texture of domain walls, observed in some non-centrosymmetric systems. Asymmetric nucleation measurements constitute a straightforward way to determine the sign of $D$ and therefore
the chirality of N\'{e}el walls.
Note that edge nucleation of reversed domains bypasses the topological problem of nucleating skyrmions inside
a sample \cite{Sampaio2013,LeePRB2014}.

A droplet model including DMI gives quantitative agreement with the measurements,
even when assuming a complete rigidity of the magnetic structures.
A full treatment of the domain wall profile shows that for large DMI and in-plane field much lower than the anisotropy field the DDW energy becomes
negative, a feature hitherto unnoticed.
Although all consequences of this specific feature of the Dzyaloshinskii domain walls need to be explored, we
already observed nucleation of reversed domains under the application of the sole $H_x$ field.

This work was supported by the Agence Nationale de la Recherche, project ANR 11 BS10 008 ESPERADO and by the Nanoscience Foundation, project MIDWEST.
SR and AT thank Jacques Miltat for discussions. SP acknowledges the support of E. Wagner and of the Nanofab team at the Institut N\'{e}el.

\newpage
\part{
Supplemental Materials}

\section{EXPERIMENTAL SETUP AND PROCEDURE}

\begin{figure}[ht!]
\includegraphics[width=8cm]{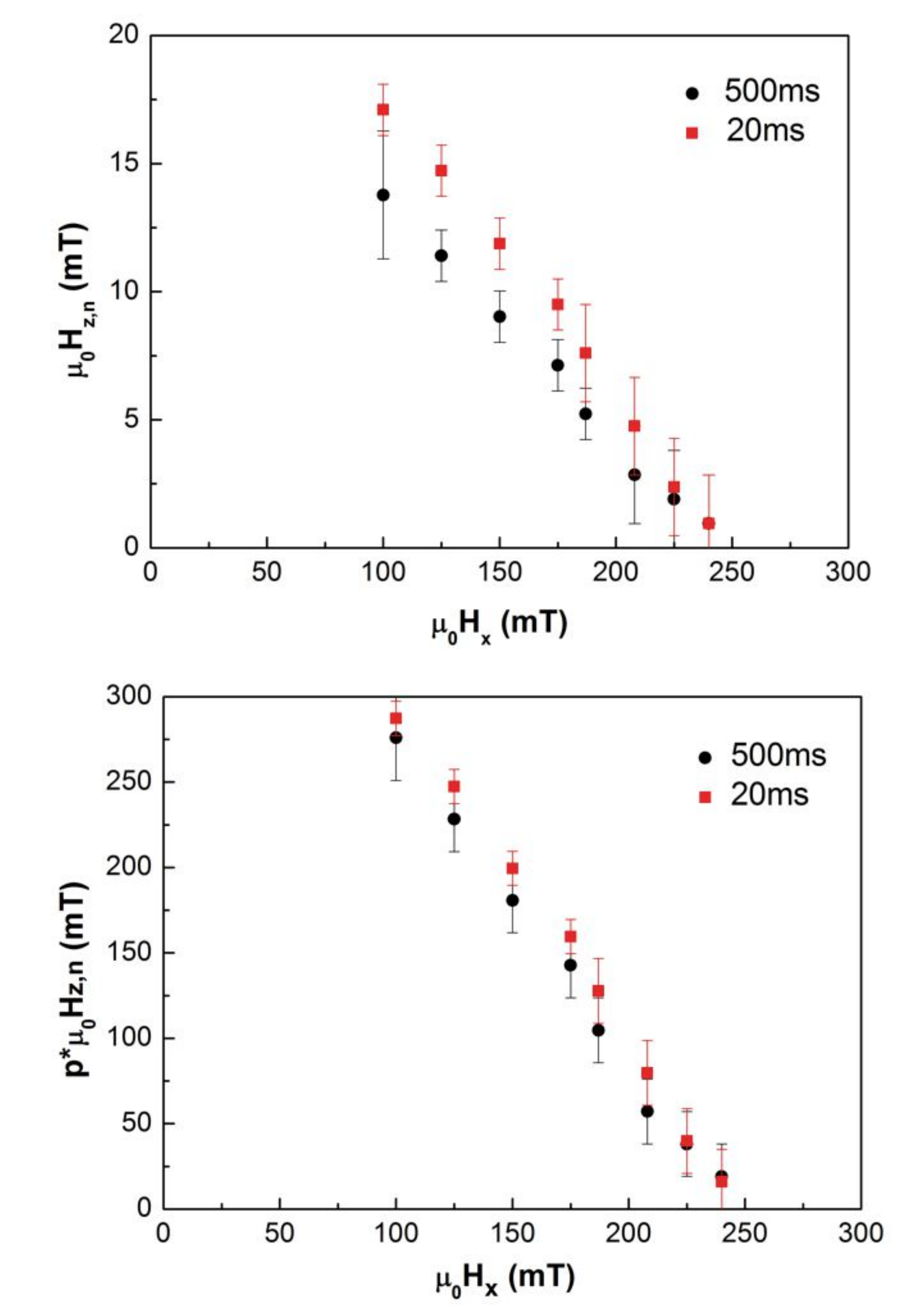}
\caption{\label{fig:Suppl-Figure1-experiment} (a) nucleation fields $H_{z,n}$ measured as a function of $H_{x}$ for two values
of the applied $H_{z}$ pulse length, 20~ms and 500~ms.
(b) the data in (a) have been multiplied by p, respectively 16.8 and 20.  }
\end{figure}

Measurements were carried out using a wide-field Magneto-Optical Kerr Effect microscope in polar geometry using a 630~nm LED.
Figure~\ref{fig:Suppl-Figure1-experiment}(a) shows the variation of the nucleation field $H_{z,n}$ at the edge of the sample, measured for
two values of the applied $H_{z}$ pulse length, $20$~ms and $500$~ms.
For a fixed $H_{x}$ field, we have (arbitrarily) defined as the \textit{nucleation field} the $H_{z}$ field for which 10-15 domains
nucleate systematically at the edge.
The experimental procedure was the following: with the $H_{x}$  field switched on, a $H_{z}$ pulse of fixed length was applied.
The $H_{x}$ field was then switched off, and an $H_{z}$ field pulse was applied to allow the created domain to enlarge, via domain
wall propagation, to a size larger than the spatial resolution of the Kerr microscope.
In order to account for the unavoidable misalignment of the $H_{x}$ electromagnet that can introduce a weak $H_{z}$ component adding
to the $H_{z}$ pulse,  the procedure was carried out twice, starting from positive and negative magnetic saturation, and the
presented values are the average of the two measurements.

Note that a larger error bar is associated to $H_{z,n}$ corresponding to the smallest $H_{x}$ values.
This is due to the fact that when the nucleation field for domains within the film becomes smaller than that at the edges, the
observation of domain nucleation at the edge is hindered by the expansion of the domains within the film.
The criterium which we have used to define $H_{z,n}$ is more difficult to obtain, making the estimation of this field subject
to larger errors.
It should be noted that several samples with the same magnetic properties have been measured using the same experimental procedure.
The minimum field for which the $H_{z,n}$ field is obtained depends on the number and position of the defects within the film.

We have shown in the manuscript that the nucleation field depends on the length of the $H_{z}$ pulse through the coefficient
$1/p$ where $p=\ln(\tau/\tau_{0})$.
For the two $H_{z}$ pulses of 20~ms and 500~ms, $p$ is respectively 16.8 and 20 when assuming $\tau_{0}=1$~ns, as usually found
in the literature.
Figure~\ref{fig:Suppl-Figure1-experiment}(b) shows the $H_{z,n}$  values multiplied by the respective $p$ values.
The invariance of the results with respect to  $p$, within the experimental error bars, is in agreement with the theoretical model.

\section{ZERO TEMPERATURE MODEL}

We consider a semi-infinite 1D model to describe the magnetization profile in the vicinity of an edge of the sample.
Let $x$ be the coordinate normal to the edge, and $z$ the easy magnetization axis.
As the field is applied in the ($x, z$) plane, and as DMI forces a magnetization rotation in this plane too, the single unknown
is the angle $\theta(x)$ of the magnetization with the easy axis.
The energy density reads then
\begin{equation}\label{energy}
E = A\left(\frac{d\theta}{dx}\right)^2 - D\frac{d\theta}{dx} + K_0 \sin^2\theta - \mu_0M_s\left(H_x\sin\theta + H_z\cos\theta\right)
\end{equation}
where $K_0 = K_u - \mu_0 M_s^2/2$ is the effective anisotropy constant including the shape anisotropy.
Writing down the Lagrange-Euler equations for the minimization of the total energy and integrating back gives the following first integral
\begin{equation}\label{first_integr}
A\left(\frac{d\theta}{dx}\right)^2 = K_0 \sin^2\theta - \mu_0M_s\left(H_x\sin\theta + H_z\cos\theta\right) + C^{st}
\end{equation}

Infinitely inside the sample, the angle is constant ($\theta_0$ in the following) so that the energy reduces to a macrospin one:
\begin{equation}\label{macrospin}
K_0e(\theta) = K_0 \sin^2\theta - \mu_0M_s(H_x\sin\theta + H_z\cos\theta)
\end{equation}
The angle $\theta_0$ minimizes this energy, which is solved by the standard Stoner-Wohlfarth problem under a general
 field \cite{Stoner1948SI,Thiaville1998SI}.

Thus, Eq.~(\ref{first_integr}) can be re-expressed as
\begin{equation}\label{etheta}
A\left(\frac{d\theta}{dx}\right)^2 = K_0\left[e(\theta) - e(\theta_0)\right]
\end{equation}
the right-hand side being indeed positive as $\theta_0$ minimizes $e(\theta)$.

At the edge of the sample, the micromagnetic boundary condition with DMI reads \cite{Rohart2013}
\begin{equation}\label{boundary}
2A\frac{d\theta}{dx} = D
\end{equation}
Let us call $\theta_1$ the magnetization angle at the edge.
From Eq.~(\ref{etheta}) and Eq.~(\ref{boundary}) we have
\begin{equation}\label{E_DMI}
e(\theta_1) - e(\theta_0) = \frac{D^2}{4AK_0} \equiv \left(\frac{2}{\pi}\frac{D}{D_{c0}}\right)^2
\end{equation}
This shows that the problem reduces to a two-macrospin problem.
The first macrospin $\theta_0$ is independently solved and an analytical solution can be found \cite{Stoner1948SI,Thiaville1998SI}.
Concerning the second macrospin (edge magnetization), Eq.~(\ref{E_DMI}) shows that its energy is larger than the previous one which lets
expect a switching at smaller fields than for $\theta_0$, or in other words, an easier magnetization switching at the edges than inside the film.

\begin{figure}
\includegraphics[width=\columnwidth]{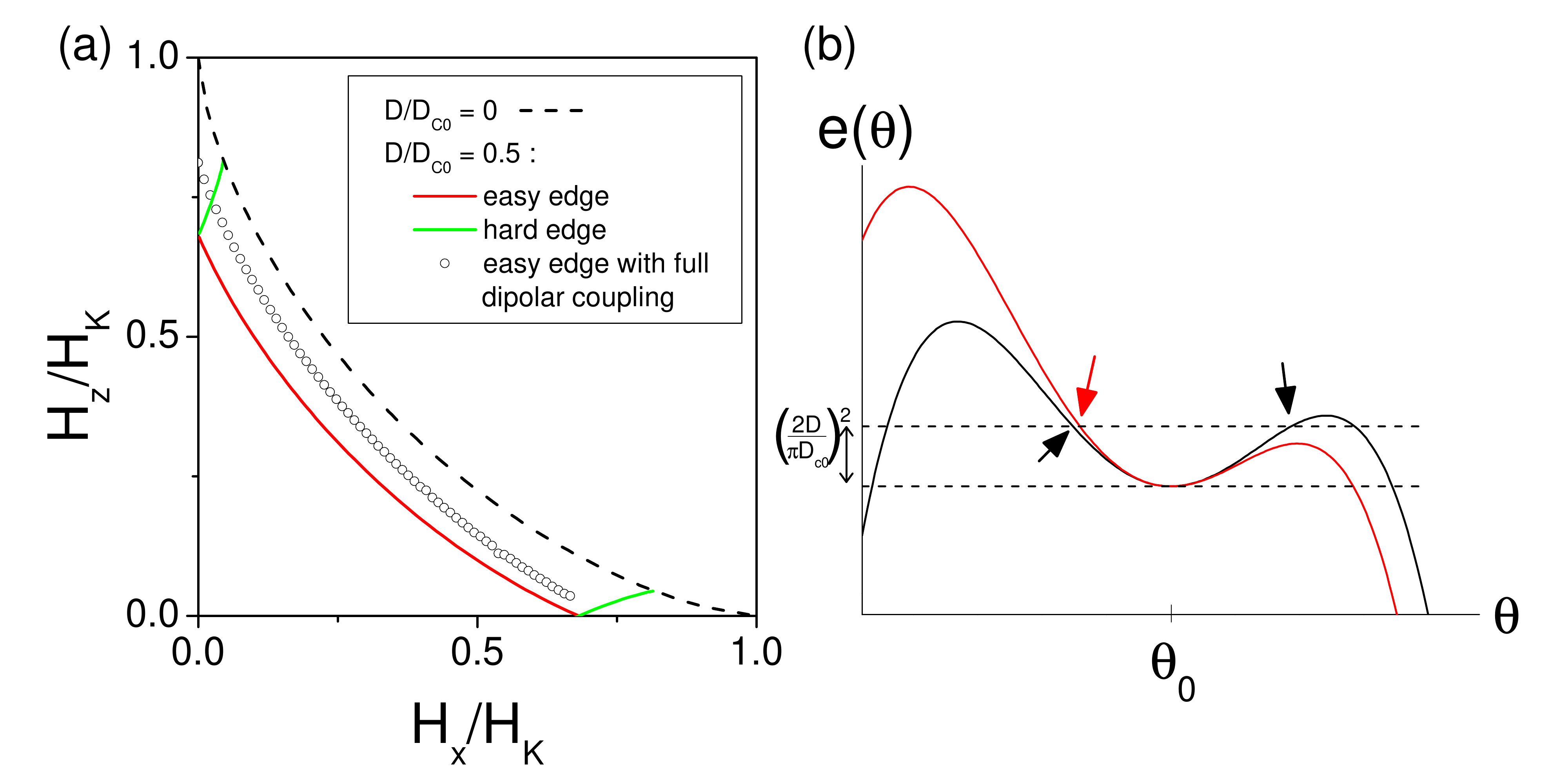}
\caption{\label{fig:Suppl-model-SW2} (a) Nucleation field $H_z$ versus $H_x$ at the edges of the sample, obtained for $D=0$ and $D = D_{c0}/2$
from Eq.~\ref{E_DMI}.
For $D\neq0$ two branches are obtained, depending on the considered edge orientation.
The dots correspond to the micromagnetic simulation including complete dipolar coupling (it has been checked that, in the local
demagnetizing field approximation, simulations are in perfect agreement with the model).
(b) Sketch of the calculation using Eq.~\ref{E_DMI} for two different field magnitudes, and for a DMI energy scale indicated by the
distance between the two dotted lines.
The solutions for $\theta_1$ are the intersection between the upper dotted line and the full curves (indicated by the arrows).
For the lower field magnitude (in black), two solutions for $\theta_1$ are found corresponding to both edges.
For the higher one (in red), one solution disappears indicating that one edge magnetization is unstable and that nucleation occurs.}
\end{figure}

Around the metastable solution for $\theta_0$, two solutions for $\theta_1$ can be found which correspond to the two edges,
$\theta_1-\theta_0$ being of the sign of $D$ on the right edge (see Eq.~\ref{boundary}).
If the symmetry is broken ($H$ arbitrarily oriented in space), the two solutions will disappear (which corresponds to a nucleation event) at
different applied field magnitudes, revealing the chirality of edge nucleation process.
Note that the easiest branches are iso-barrier curves of the Stoner-Wohlfarth problem. These have already been investigated, when studying
the switching field of single-domain nanoparticles at non-zero temperatures, both experimentally through single particle measurements
using the micro SQUID technique \cite{Jamet2001SI}, and theoretically in the macrospin model \cite{Vouille2004SI}.
These curves are simply shrunk astroids.
The solutions for $D/D_{c0} = 0$ and 0.5 are shown in Fig.~\ref{fig:Suppl-model-SW2}.

This model neglects dipolar couplings (only the shape anisotropy is taken into account), as generally done for ultrathin films.
However, the comparison to full micromagnetic simulations (2D mesh, modified OOMMF code) that include the demagnetizing field exactly,
is also shown in Fig.~\ref{fig:Suppl-model-SW2}.
Small differences are seen, especially for small in-plane fields because of the magnetostatic charge that appears at the sample
edge reduces the edge magnetization tilt \cite{Rohart2013a}.

\section{MICROMAGNETIC SIMULATIONS}

We consider a magnetic ultrathin film grown on a substrate with a capping layer in a different material so that the
inversion symmetry is broken along the vertical axis ($z$).
The magnetization is oriented out-of-plane with a strong perpendicular anisotropy.
In addition to the standard micromagnetic energy density which includes the exchange, the magnetocrystalline anisotropy,
the Zeeman and the demagnetizing energy, we add the Dzyaloshinskii-Moriya contribution (DM) that reads in a continuous form
\cite{Thiaville2012a}:
\begin{equation}\label{DMI}
E_{DM}=D[m_{z}\frac{\partial m_{x}}{\partial x}-m_{x}\frac{\partial m_{z}}{\partial x}+ id.(x\rightarrow y)]
\end{equation}
Micromagnetic simulations are based on the Landau-Lifshitz-Gilbert equation integration:
\begin{equation}\label{DMI2}
\frac{\partial \textbf{m}}{\partial t} = -\gamma_0[\textbf{m} \times \textbf{H$_{eff}$}]+
\alpha(\textbf{m} \times \frac{\partial \textbf{m}}{\partial \textbf{t}})
\end{equation}

3D micromagnetic simulations are performed using the homemade micromagnetic solver Micro3D \cite{Buda2002SI}.
The following material parameters have been used: exchange stiffness $A=1\times10^{-11}$~J/m,
saturation magnetization $M_{s} =1.09\times 10^{6}$~A/m, uniaxial magnetocrystalline anisotropy constant
$K = 1.25 \times 10^{6}$~J/m$^{3}$ (along the out-of-plane direction $z$), Dzyaloshinskii-Moriya constant $D =2$~mJ/m$^{2}$,
Gilbert damping parameter $\alpha=~0.5$.
They differ slightly from those considered in the main text (mainly by the lower $A$ and higher anisotropy
$H_{K0}= 0.92$~T), but correspond to a previous work \cite{Boulle2013SI}.

The thermal fluctuations were neglected and the size of the sample was 100~nm~$\times$~1024~nm $\times$~0.6~nm.
Starting from positive out-of-plane remanent state, two types of situation were analyzed:
\textit{i)} simultaneous application of an in plane field $H_{x}$ (parallel to the short edge of the sample) and an easy axis $H_{z}$
field (perpendicular to the film plane).
The Dzyaloshinskii-Moriya  constant was first set to $D=0$~J/m$^{2}$ and after to $D=2$~mJ/m$^{2}$; ii) application of one
single in-plane $H_{x}$ field with $D=2$~mJ/m$^{2}$.

\begin{figure*}[ht!]
\includegraphics[width=16cm]{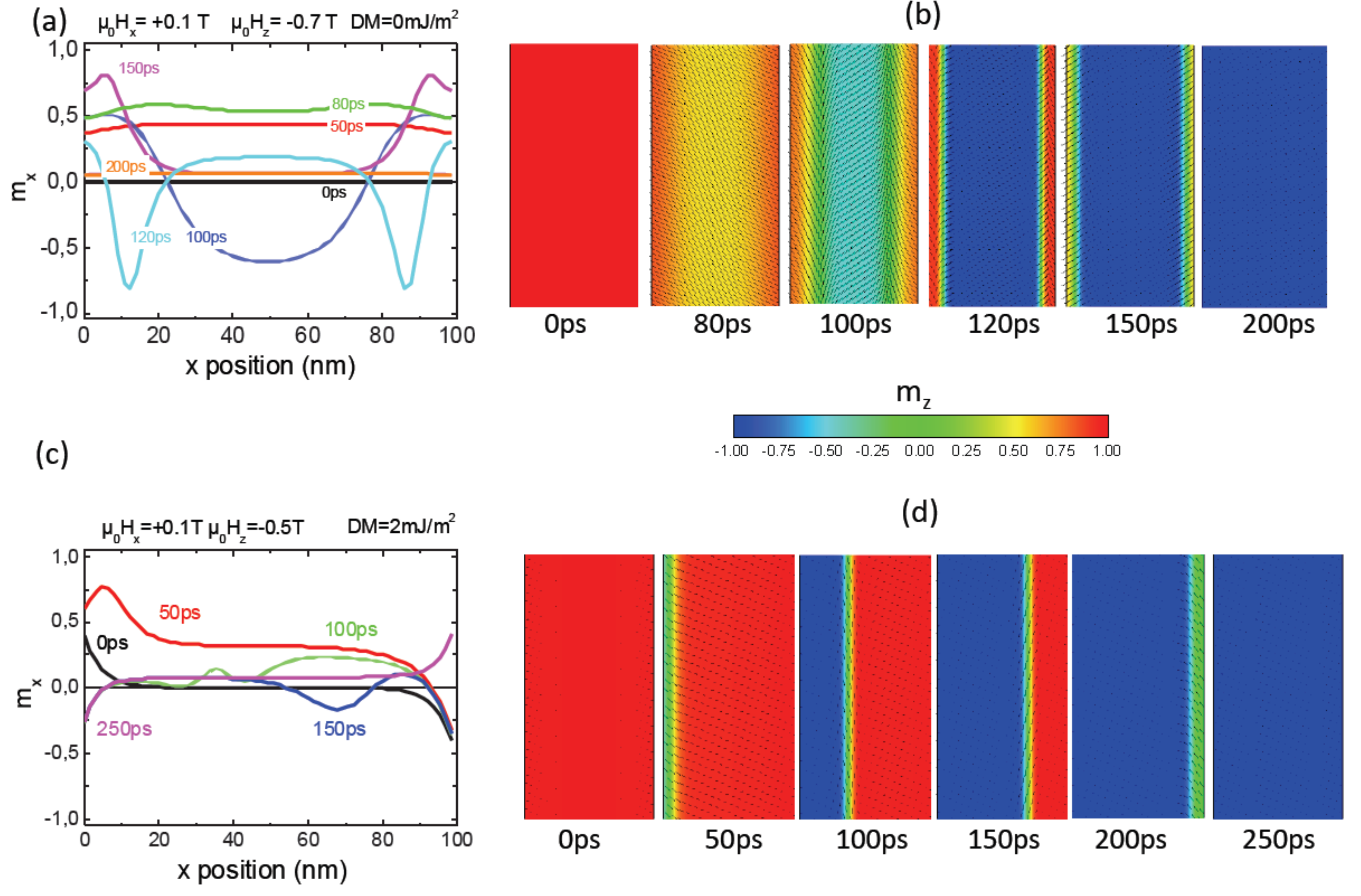}
\caption{\label{fig:Suppl-micro1}  Micromagnetic simulations showing the modification of the micromagnetic structure of
the sample under the effect of: (a) $\mu_{0}H_{z}$= -0.7~T and $\mu_{0}H_{x}$=+0.1~T  for
$D$~=~0~J/m$^{2}$ - the reversal occurs starting from the center of the sample;
(b) $\mu_{0}H_{z}$= -0.5~T and $\mu_{0}H_{x}$=+0.1~T for $D$=~2~mJ/m$^{2}$ - the reversal starts from the left
edge of the sample.}
\end{figure*}

Figure~\ref{fig:Suppl-micro1}(a-b) summarizes the results of the simulations carried out applying $\mu_{0}$H$_{x} = +0.1$~T
and $\mu_{0}$H$_{z}=-0.7$~T for $D=0$~J/m$^{2}$.
Starting from out of plane remanent state  at $t=0$~ps, the magnetization reverses by nucleation of a domain in the center of
the sample, where the demagnetizing field is the largest.
The magnetization reversal follows a symmetric mechanism.
Figure~\ref{fig:Suppl-micro1}(a) shows the time evolution of the in-plane $m_{x}$ component of the magnetization across the
short edge of the sample. Snapshots at different moments of time are shown on Fig.~\ref{fig:Suppl-micro1}(b): the color is
related to the out-of-plane magnetization and the arrows give the in-plane magnetization vector projection.
The magnetization reversal starts in the middle of the sample and propagates to the edges.

\begin{figure*}[ht!]
\includegraphics[width=16cm]{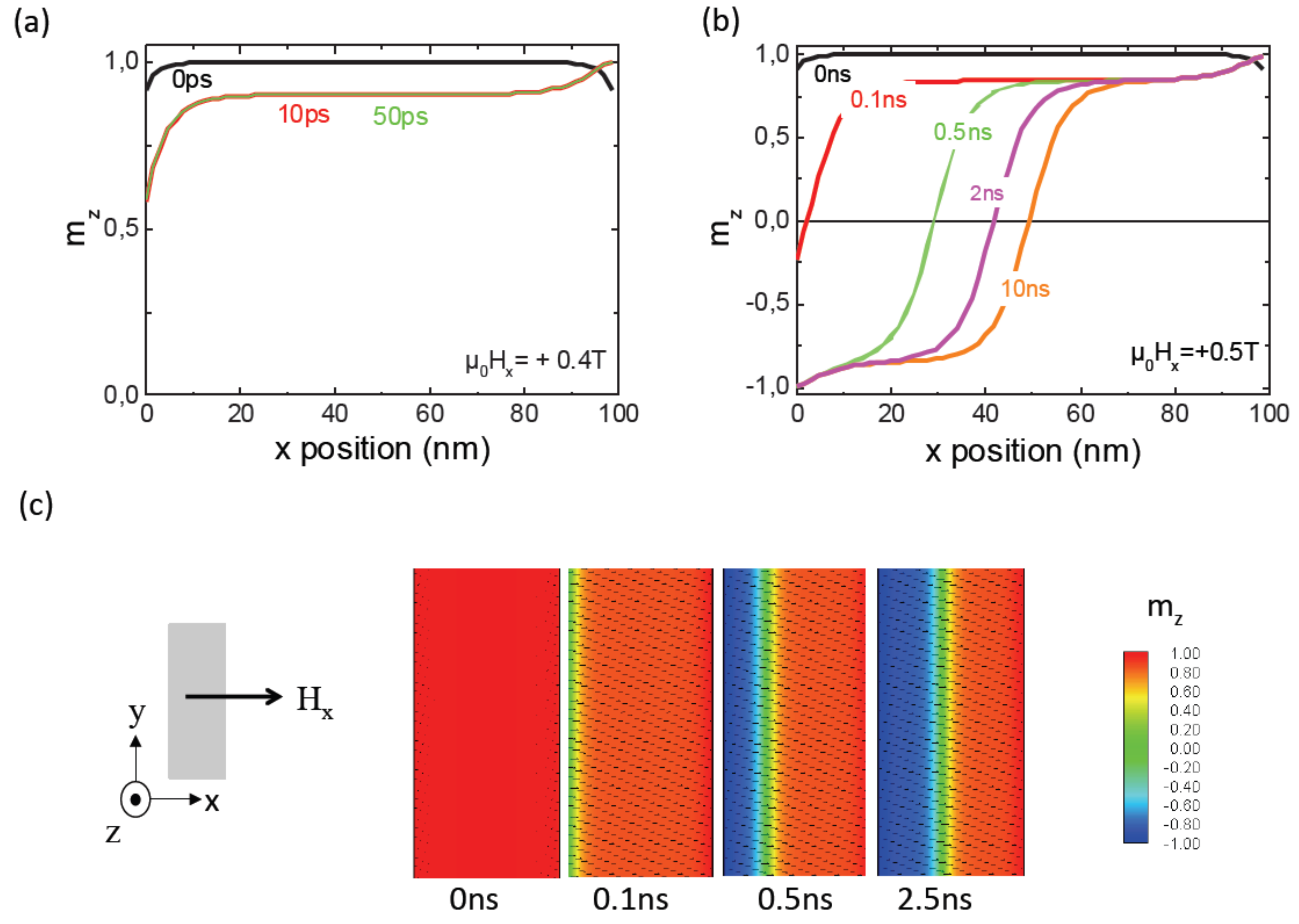}
\caption{\label{fig:Suppl-micro2} Micromagnetic simulations showing the modification of the micromagnetic structure of the
sample under the effect of an in-plane $H_{x}$ field applied at time $t= 0$~ps for $D=2$~mJ/m$^{2}$.
(a) for $\mu_{0}H_{x}=+0.4$~T the magnetization tilts in the direction of the applied field;
(b) for $\mu_{0}$H$_{x}~=+0.5$~T a domain wall is formed at the left edge of the sample and stabilizes away from the edge
due to magnetostatic effects;
(c) snapshots of the magnetization illustrating the evolution towards equilibrium of the domain wall formed under
$\mu_{0}$H$_{x}=+0.5$~T.}
\end{figure*}

Figure \ref{fig:Suppl-micro1}(c-d) shows that the magnetization reversal changes drastically in the presence of DMI.
Before the application of the field, the magnetization at the sides of the sample is tilted towards the center, in opposite
directions on either side of the sample.
The effect of the $x$-field is to tilt the magnetization in the direction of the applied field; with a positive $H_{x}$,
the $m_{x}$ component of the magnetization will have a maximum value on the left side of the sample.
Magnetization reversal is then triggered by nucleation of reversed domains at this edge, where the $m_{x}$ component of
magnetization is initially parallel to the $H_{x}$ field.

Figure~\ref{fig:Suppl-micro2} (a) shows the effect of an in-plane field $\mu_{0}H_{x}=+0.4$~T applied at time $t=0$~ps,
for the case where $=2$~mJ/m$^{2}$.
For this value of the $x$-field,  the magnetization simply tilts in the direction of the applied field.
The situation is very different for a larger $H_{x}$ field of +0.5~T (Fig.~\ref{fig:Suppl-micro2}(b)).
In this case, a domain wall is created starting from the left edge of the sample, and its position stabilizes around the center
of the sample due to demagnetizing effects.
The creation of this domain wall in the presence of an in-plane field agrees with the results of the model developed in the main
text of this Letter, which shows that in the presence of a large DMI, the domain wall energy becomes negative beyond a
threshold in-plane field, therefore favoring its formation.
Figure~\ref{fig:Suppl-micro2}(c) gives a 2D representation of the dynamics of the magnetization and its evolution towards
equilibrium, showing the initial tilt of the magnetization at one edge of the sample, the formation of the domain wall and its
propagation away from the edge.

\section{ESTIMATION OF THE MAGNETIC PARAMETERS OF THE SAMPLE}

The various micromagnetic parameters of the sample are not exactly known: some of them are directly obtained from
characterizations, but some others are more difficult to measure, so that one has to see whether or not, with reasonable
values assumed for them, the experimental results can be reproduced.
In the case at hand, the relevant parameters are: spontaneous magnetization $M_s$,
exchange constant $A$,
effective anisotropy $K_0$ (the sum of the volumic, interfacial and shape anisotropies),
thickness averaged interfacial DMI $D$, and $p$ the
number of thermal energy quanta $k_B T$ involved in the Arrhenius process, given by
$\tau = \tau_0$e$^p$ with $\tau_0$ the attempt time and $\tau$ the duration of field application for the domain nucleation.
We assume for the calculations that $\tau_0 = 1$~ns.
The value 0.1~ns is often quoted but, as in the Brown theory the attempt time is linked to the damping constant $\alpha$,
and because from the DW dynamics experiments \cite{Metaxas2007SI} a value of $\alpha$ about 10 times larger than for bulk cobalt
was deduced, we took this larger value.
Note also that taking $\tau_0 = 0.1$~ns increases $p$ by 2.3 only, for a typical value of 18 that corresponds to $\tau = 66$~ms.
Finally, the DW energy reduction at the defect in the center of the film is given by the factor $\epsilon$.

\begin{figure*}[ht!]
\includegraphics[width=16cm]{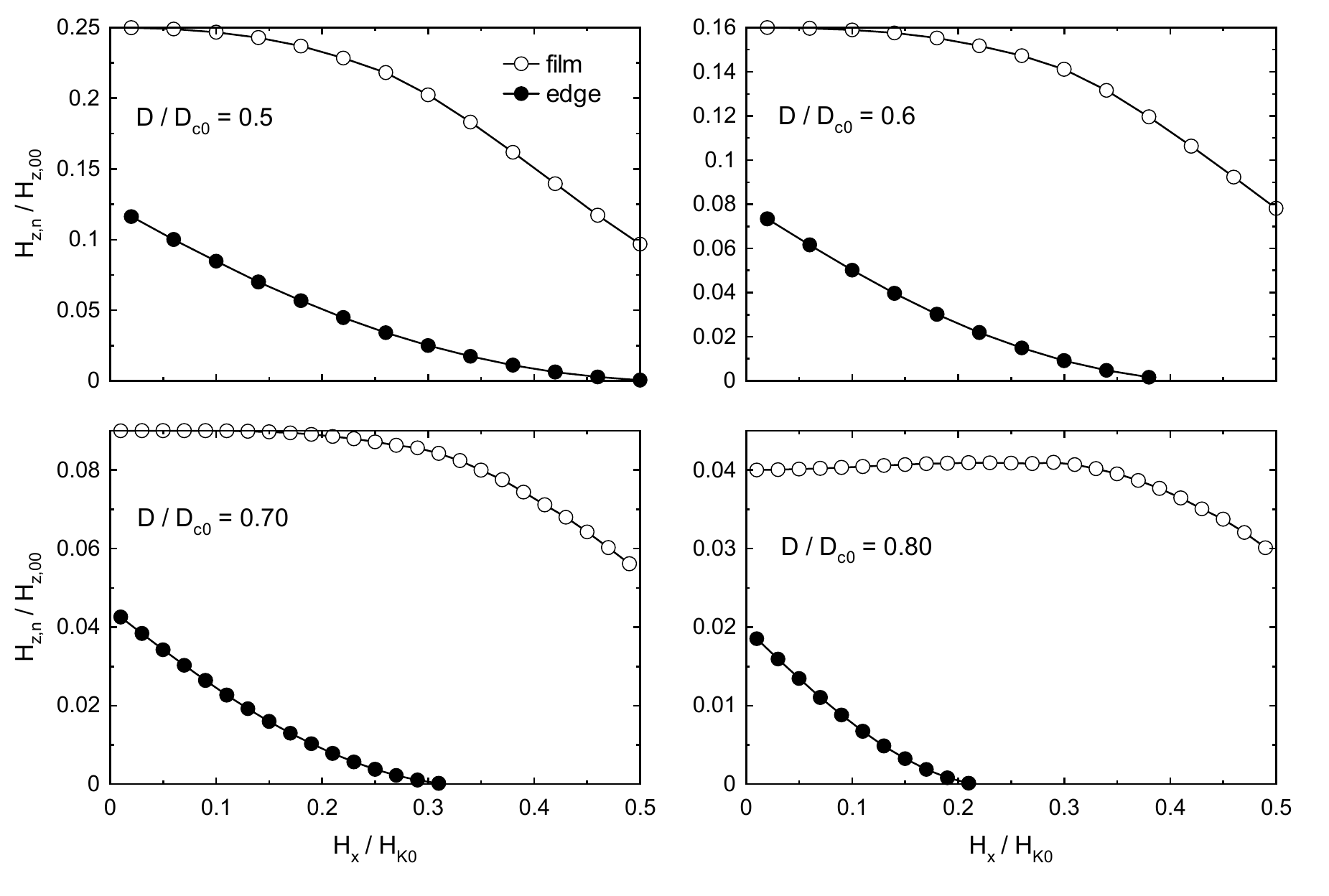}
\caption{\label{fig:Suppl-FigureC1} Nucleation fields in the film and at the edge, normalized to $H_{z,00}$ (see Eq.~(\ref{eq:Hz00})),
as a function of $x$ field normalized to the effective anisotropy field $H_{K0}$, for different values of the normalized DMI,
as computed by the semi-analytical model.
The horizontal scale is always the same, but note the change of the vertical scale as DMI varies.}
\end{figure*}

For $M_s$, we have taken the value 1.1~MA/m.
The bulk Co value is 1.4~MA/m.
SQUID measurements, with large error bars as the sample thickness is only 3 atomic layers ($t= 0.6$~nm), are consistent with
a lower value of 1.1~MA/m, in agreement with literature.

The effective anisotropy field $H_{K0} = 2K_0 / (\mu_0 M_s)$ was directly measured by extraordinary Hall effect under
in-plane field. We found $\mu_0 H_{K0} = 0.7$~T.

The exchange constant was assumed to amount to 16~pJ/m.
For bulk Co, the literature contains values ranging from 10 to 30~pJ/m.
The value used here corresponds to an estimate based on the Curie temperature of the ultrathin films \cite{Metaxas2007SI}.
We will discuss below what changes if a lower value of 9~pJ/m is taken.

With these parameters, the primitive DW energy density (without DMI, and under no applied in-plane field) is
$\sigma_{00}$ = 4 ($AK_0$)$^{1/2}$ = 9.92~mJ/m$^2$, thus the critical DMI for destabilization of the uniform state is
$D_{c0}$ = $\sigma_{00}/\pi$ = 3.16~mJ/m$^2$.
The DW width parameter is $\Delta = (A/K_0)^{1/2} = 6.4$~ nm.
From these values, we can evaluate the scaling $z$ field of the problem, given by

\begin{equation}
\label{eq:Hz00}
H_{z00} = \frac{\pi \sigma_{00}^2 t}{2\mu_0M_spk_BT} = \frac{4\pi t}{pk_BT}AH_{K0}
\end{equation}
and find $\mu_0 H_{z00} = 1172$~mT.

In order to estimate the value of the DMI parameter $D$, we have systematically varied the $D/D_{c0}$ ratio and computed
the nucleation fields in the film and at the edge, keeping $\epsilon = 1$ for the moment so that the film nucleation field
is very large.
Results are shown in Fig.~\ref{fig:Suppl-FigureC1}.
From both the value of the in-plane field where the edge nucleation field becomes zero (about 250~mT i.e. $H/H_{K0}=0.36$),
and the shape of the curve showing the nucleation field in the film, we find that $D/D_{c0} = 0.7$ fits the best.

Finally, in order to reproduce the experimental value of the nucleation field in the film, we need $\epsilon= 0.42$.
From Eq.~(\ref{eq:Hz00}), we see that the value of the nucleation field in the film will be given by the product $A\epsilon^2$.
Thus, with a lower value $A=9$~pJ/m, a value $\epsilon = 0.57$ would be required.
This would however shift the value of the in-plane field where the two nucleation fields are equal, from 105 to 56~mT,
whereas experimentally it is about 125~mT.

A word of caution is in order here about the physical meaning of this defect parameter $\epsilon$.
It was originally introduced in order to model the effect of the defect on the nucleation field \cite{Moritz2005a}, in the
case of a sole easy axis field, through a reduction of the DW energy.
Here, when applying an in-plane field, the DW energy becomes orientation dependent so that the validity of an isotropic
reduction of its value by the factor $\epsilon$ should be investigated in more detail.
We leave this as a subject of future work.


\begin{thebibliography}{0}%
\makeatletter
\providecommand \@ifxundefined [1]{%
 \@ifx{#1\undefined}
}%
\providecommand \@ifnum [1]{%
 \ifnum #1\expandafter \@firstoftwo
 \else \expandafter \@secondoftwo
 \fi
}%
\providecommand \@ifx [1]{%
 \ifx #1\expandafter \@firstoftwo
 \else \expandafter \@secondoftwo
 \fi
}%
\providecommand \natexlab [1]{#1}%
\providecommand \enquote  [1]{``#1''}%
\providecommand \bibnamefont  [1]{#1}%
\providecommand \bibfnamefont [1]{#1}%
\providecommand \citenamefont [1]{#1}%
\providecommand \href@noop [0]{\@secondoftwo}%
\providecommand \href [0]{\begingroup \@sanitize@url \@href}%
\providecommand \@href[1]{\@@startlink{#1}\@@href}%
\providecommand \@@href[1]{\endgroup#1\@@endlink}%
\providecommand \@sanitize@url [0]{\catcode `\\12\catcode `\$12\catcode
  `\&12\catcode `\#12\catcode `\^12\catcode `\_12\catcode `\%12\relax}%
\providecommand \@@startlink[1]{}%
\providecommand \@@endlink[0]{}%
\providecommand \url  [0]{\begingroup\@sanitize@url \@url }%
\providecommand \@url [1]{\endgroup\@href {#1}{\urlprefix }}%
\providecommand \urlprefix  [0]{URL }%
\providecommand \Eprint [0]{\href }%
\providecommand \doibase [0]{http://dx.doi.org/}%
\providecommand \selectlanguage [0]{\@gobble}%
\providecommand \bibinfo  [0]{\@secondoftwo}%
\providecommand \bibfield  [0]{\@secondoftwo}%
\providecommand \translation [1]{[#1]}%
\providecommand \BibitemOpen [0]{}%
\providecommand \bibitemStop [0]{}%
\providecommand \bibitemNoStop [0]{.\EOS\space}%
\providecommand \EOS [0]{\spacefactor3000\relax}%
\providecommand \BibitemShut  [1]{\csname bibitem#1\endcsname}%
\let\auto@bib@innerbib\@empty
\end{thebibliography}%


\begin{thebibliography}{31}%
\makeatletter
\providecommand \@ifxundefined [1]{%
 \@ifx{#1\undefined}
}%
\providecommand \@ifnum [1]{%
 \ifnum #1\expandafter \@firstoftwo
 \else \expandafter \@secondoftwo
 \fi
}%
\providecommand \@ifx [1]{%
 \ifx #1\expandafter \@firstoftwo
 \else \expandafter \@secondoftwo
 \fi
}%
\providecommand \natexlab [1]{#1}%
\providecommand \enquote  [1]{``#1''}%
\providecommand \bibnamefont  [1]{#1}%
\providecommand \bibfnamefont [1]{#1}%
\providecommand \citenamefont [1]{#1}%
\providecommand \href@noop [0]{\@secondoftwo}%
\providecommand \href [0]{\begingroup \@sanitize@url \@href}%
\providecommand \@href[1]{\@@startlink{#1}\@@href}%
\providecommand \@@href[1]{\endgroup#1\@@endlink}%
\providecommand \@sanitize@url [0]{\catcode `\\12\catcode `\$12\catcode
  `\&12\catcode `\#12\catcode `\^12\catcode `\_12\catcode `\%12\relax}%
\providecommand \@@startlink[1]{}%
\providecommand \@@endlink[0]{}%
\providecommand \url  [0]{\begingroup\@sanitize@url \@url }%
\providecommand \@url [1]{\endgroup\@href {#1}{\urlprefix }}%
\providecommand \urlprefix  [0]{URL }%
\providecommand \Eprint [0]{\href }%
\providecommand \doibase [0]{http://dx.doi.org/}%
\providecommand \selectlanguage [0]{\@gobble}%
\providecommand \bibinfo  [0]{\@secondoftwo}%
\providecommand \bibfield  [0]{\@secondoftwo}%
\providecommand \translation [1]{[#1]}%
\providecommand \BibitemOpen [0]{}%
\providecommand \bibitemStop [0]{}%
\providecommand \bibitemNoStop [0]{.\EOS\space}%
\providecommand \EOS [0]{\spacefactor3000\relax}%
\providecommand \BibitemShut  [1]{\csname bibitem#1\endcsname}%
\let\auto@bib@innerbib\@empty
\bibitem [{\citenamefont {Mislow}(1999)}]{Mislow1999}%
  \BibitemOpen
  \bibfield  {author} {\bibinfo {author} {\bibfnamefont {K.}~\bibnamefont
  {Mislow}},\ }\enquote {\bibinfo {title} {Molecular chirality},}\ \ (\bibinfo
  {publisher} {Wiley},\ \bibinfo {address} {New York},\ \bibinfo {year}
  {1999})\ pp.\ \bibinfo {pages} {1--82},\ \bibinfo {note} {topics in
  stereochemistry, vol. 22, S.E. Denmark Ed.}\BibitemShut {Stop}%
\bibitem [{not({\natexlab{a}})}]{note-chiralite}%
  \BibitemOpen
  \href@noop {} {}\bibinfo {note} {The word chirality was
  introduced later by Lord Kelvin, in 1893 [1].}\BibitemShut
  {Stop}%
\bibitem [{\citenamefont {Volovik}\ and\ \citenamefont
  {Krusius}(2012)}]{Volovik2012}%
  \BibitemOpen
  \bibfield  {author} {\bibinfo {author} {\bibfnamefont {G.~E.}\ \bibnamefont
  {Volovik}}\ and\ \bibinfo {author} {\bibfnamefont {M.}~\bibnamefont
  {Krusius}},\ }\href@noop {} {\bibfield  {journal} {\bibinfo  {journal}
  {Physics}\ }\textbf {\bibinfo {volume} {5}},\ \bibinfo {pages} {130}
  (\bibinfo {year} {2012})}\BibitemShut {NoStop}%
\bibitem [{\citenamefont {de~Gennes}\ and\ \citenamefont
  {Prost}(1993)}]{deGennes1993}%
  \BibitemOpen
  \bibfield  {author} {\bibinfo {author} {\bibfnamefont {P.-G.}\ \bibnamefont
  {de~Gennes}}\ and\ \bibinfo {author} {\bibfnamefont {J.}~\bibnamefont
  {Prost}},\ }\href@noop {} {\emph {\bibinfo {title} {The physics of liquid
  crystals, 2nd. Ed.}}}\ (\bibinfo  {publisher} {Oxford University Press},\
  \bibinfo {year} {1993})\BibitemShut {NoStop}%
\bibitem [{\citenamefont {Dzyaloshinskii}(1957)}]{Dzyaloshinskii1957}%
  \BibitemOpen
  \bibfield  {author} {\bibinfo {author} {\bibfnamefont {I.~E.}\ \bibnamefont
  {Dzyaloshinskii}},\ }\href@noop {} {\bibfield  {journal} {\bibinfo  {journal}
  {Sov. Phys. JETP}\ }\textbf {\bibinfo {volume} {5}},\ \bibinfo {pages} {1259}
  (\bibinfo {year} {1957})}\BibitemShut {NoStop}%
\bibitem [{\citenamefont {Moriya}(1960)}]{Moriya1960}%
  \BibitemOpen
  \bibfield  {author} {\bibinfo {author} {\bibfnamefont {T.}~\bibnamefont
  {Moriya}},\ }\href@noop {} {\bibfield  {journal} {\bibinfo  {journal} {Phys.
  Rev.}\ }\textbf {\bibinfo {volume} {120}},\ \bibinfo {pages} {91} (\bibinfo
  {year} {1960})}\BibitemShut {NoStop}%
\bibitem [{\citenamefont {M{\"{u}}hlbauer}\ \emph {et~al.}(2009)\citenamefont
  {M{\"{u}}hlbauer}, \citenamefont {Binz}, \citenamefont {Jonietz},
  \citenamefont {Pfleiderer}, \citenamefont {Rosch}, \citenamefont {Neubauer},
  \citenamefont {Georgii},\ and\ \citenamefont {B{\"{o}}ni}}]{Muhlbauer2009}%
  \BibitemOpen
  \bibfield  {author} {\bibinfo {author} {\bibfnamefont {S.}~\bibnamefont
  {M{\"{u}}hlbauer}}, \bibinfo {author} {\bibfnamefont {B.}~\bibnamefont
  {Binz}}, \bibinfo {author} {\bibfnamefont {F.}~\bibnamefont {Jonietz}},
  \bibinfo {author} {\bibfnamefont {C.}~\bibnamefont {Pfleiderer}}, \bibinfo
  {author} {\bibfnamefont {A.}~\bibnamefont {Rosch}}, \bibinfo {author}
  {\bibfnamefont {A.}~\bibnamefont {Neubauer}}, \bibinfo {author}
  {\bibfnamefont {R.}~\bibnamefont {Georgii}}, \ and\ \bibinfo {author}
  {\bibfnamefont {P.}~\bibnamefont {B{\"{o}}ni}},\ }\href@noop {} {\bibfield
  {journal} {\bibinfo  {journal} {Science}\ }\textbf {\bibinfo {volume}
  {323}},\ \bibinfo {pages} {915} (\bibinfo {year} {2009})}\BibitemShut
  {NoStop}%
\bibitem [{\citenamefont {Yu}\ \emph {et~al.}(2010)\citenamefont {Yu},
  \citenamefont {Onose}, \citenamefont {Park}, \citenamefont {Han},
  \citenamefont {Matsui}, \citenamefont {Nagaosa},\ and\ \citenamefont
  {Tokura}}]{Yu2010}%
  \BibitemOpen
  \bibfield  {author} {\bibinfo {author} {\bibfnamefont {X.~Z.}\ \bibnamefont
  {Yu}}, \bibinfo {author} {\bibfnamefont {Y.}~\bibnamefont {Onose}}, \bibinfo
  {author} {\bibfnamefont {J.~H.}\ \bibnamefont {Park}}, \bibinfo {author}
  {\bibfnamefont {J.~H.}\ \bibnamefont {Han}}, \bibinfo {author} {\bibfnamefont
  {Y.}~\bibnamefont {Matsui}}, \bibinfo {author} {\bibfnamefont
  {N.}~\bibnamefont {Nagaosa}}, \ and\ \bibinfo {author} {\bibfnamefont
  {Y.}~\bibnamefont {Tokura}},\ }\href@noop {} {\bibfield  {journal} {\bibinfo
  {journal} {Nature}\ }\textbf {\bibinfo {volume} {465}},\ \bibinfo {pages}
  {901} (\bibinfo {year} {2010})}\BibitemShut {NoStop}%
\bibitem [{\citenamefont {Bode}\ \emph {et~al.}(2007)\citenamefont {Bode},
  \citenamefont {Heide}, \citenamefont {von Bergmann}, \citenamefont
  {Ferriani}, \citenamefont {Heinze}, \citenamefont {Bihlmayer}, \citenamefont
  {Kubetzka}, \citenamefont {Pietzsch}, \citenamefont {Bl{\"{u}}gel},\ and\
  \citenamefont {Wiesendanger}}]{Bode2007}%
  \BibitemOpen
  \bibfield  {author} {\bibinfo {author} {\bibfnamefont {M.}~\bibnamefont
  {Bode}}, \bibinfo {author} {\bibfnamefont {M.}~\bibnamefont {Heide}},
  \bibinfo {author} {\bibfnamefont {K.}~\bibnamefont {von Bergmann}}, \bibinfo
  {author} {\bibfnamefont {P.}~\bibnamefont {Ferriani}}, \bibinfo {author}
  {\bibfnamefont {S.}~\bibnamefont {Heinze}}, \bibinfo {author} {\bibfnamefont
  {G.}~\bibnamefont {Bihlmayer}}, \bibinfo {author} {\bibfnamefont
  {A.}~\bibnamefont {Kubetzka}}, \bibinfo {author} {\bibfnamefont
  {O.}~\bibnamefont {Pietzsch}}, \bibinfo {author} {\bibfnamefont
  {S.}~\bibnamefont {Bl{\"{u}}gel}}, \ and\ \bibinfo {author} {\bibfnamefont
  {R.}~\bibnamefont {Wiesendanger}},\ }\href {\doibase {10.1038/nature05802}}
  {\bibfield  {journal} {\bibinfo  {journal} {Nature}\ }\textbf {\bibinfo
  {volume} {447}},\ \bibinfo {pages} {190} (\bibinfo {year}
  {2007})}\BibitemShut {NoStop}%
\bibitem [{\citenamefont {Meckler}\ \emph {et~al.}(2009)\citenamefont
  {Meckler}, \citenamefont {Mikuszeit}, \citenamefont {Pressler},
  \citenamefont {Vedmedenko}, \citenamefont {Pietzsch},\ and\ \citenamefont
  {Wiesendanger}}]{Meckler2009}%
  \BibitemOpen
  \bibfield  {author} {\bibinfo {author} {\bibfnamefont {S.}~\bibnamefont
  {Meckler}}, \bibinfo {author} {\bibfnamefont {N.}~\bibnamefont {Mikuszeit}},
  \bibinfo {author} {\bibfnamefont {A.}~\bibnamefont {Pressler}}, \bibinfo
  {author} {\bibfnamefont {E.~Y.}\ \bibnamefont {Vedmedenko}}, \bibinfo
  {author} {\bibfnamefont {O.}~\bibnamefont {Pietzsch}}, \ and\ \bibinfo
  {author} {\bibfnamefont {R.}~\bibnamefont {Wiesendanger}},\ }\href {\doibase
  10.1103/PhysRevLett.103.157201} {\bibfield  {journal} {\bibinfo  {journal}
  {Phys. Rev. Lett.}\ }\textbf {\bibinfo {volume} {103}},\ \bibinfo {pages}
  {157201} (\bibinfo {year} {2009})}\BibitemShut {NoStop}%
\bibitem [{\citenamefont {Miron}\ \emph {et~al.}(2010)\citenamefont {Miron},
  \citenamefont {Gaudin}, \citenamefont {Auffret}, \citenamefont {Rodmacq},
  \citenamefont {Schuhl}, \citenamefont {Pizzini}, \citenamefont {Vogel},\ and\
  \citenamefont {Gambardella}}]{Miron2010}%
  \BibitemOpen
  \bibfield  {author} {\bibinfo {author} {\bibfnamefont {I.~M.}\ \bibnamefont
  {Miron}}, \bibinfo {author} {\bibfnamefont {G.}~\bibnamefont {Gaudin}},
  \bibinfo {author} {\bibfnamefont {S.}~\bibnamefont {Auffret}}, \bibinfo
  {author} {\bibfnamefont {B.}~\bibnamefont {Rodmacq}}, \bibinfo {author}
  {\bibfnamefont {A.}~\bibnamefont {Schuhl}}, \bibinfo {author} {\bibfnamefont
  {S.}~\bibnamefont {Pizzini}}, \bibinfo {author} {\bibfnamefont
  {J.}~\bibnamefont {Vogel}}, \ and\ \bibinfo {author} {\bibfnamefont
  {P.}~\bibnamefont {Gambardella}},\ }\href {\doibase {10.1038/NMAT2613}}
  {\bibfield  {journal} {\bibinfo  {journal} {Nature Mater.}\ }\textbf
  {\bibinfo {volume} {9}},\ \bibinfo {pages} {230} (\bibinfo {year}
  {2010})}\BibitemShut {NoStop}%
\bibitem [{\citenamefont {Miron}\ \emph {et~al.}(2011)\citenamefont {Miron},
  \citenamefont {Moore}, \citenamefont {Szambolics}, \citenamefont
  {Buda-Prejbeanu}, \citenamefont {Auffret}, \citenamefont {Rodmacq},
  \citenamefont {Pizzini}, \citenamefont {Vogel}, \citenamefont {Bonfim},
  \citenamefont {Schuhl},\ and\ \citenamefont {Gaudin}}]{Miron2011}%
  \BibitemOpen
  \bibfield  {author} {\bibinfo {author} {\bibfnamefont {I.~M.}\ \bibnamefont
  {Miron}}, \bibinfo {author} {\bibfnamefont {T.}~\bibnamefont {Moore}},
  \bibinfo {author} {\bibfnamefont {H.}~\bibnamefont {Szambolics}}, \bibinfo
  {author} {\bibfnamefont {L.~D.}\ \bibnamefont {Buda-Prejbeanu}}, \bibinfo
  {author} {\bibfnamefont {S.}~\bibnamefont {Auffret}}, \bibinfo {author}
  {\bibfnamefont {B.}~\bibnamefont {Rodmacq}}, \bibinfo {author} {\bibfnamefont
  {S.}~\bibnamefont {Pizzini}}, \bibinfo {author} {\bibfnamefont
  {J.}~\bibnamefont {Vogel}}, \bibinfo {author} {\bibfnamefont
  {M.}~\bibnamefont {Bonfim}}, \bibinfo {author} {\bibfnamefont
  {A.}~\bibnamefont {Schuhl}}, \ and\ \bibinfo {author} {\bibfnamefont
  {G.}~\bibnamefont {Gaudin}},\ }\href {\doibase {10.1038/NMAT3020}} {\bibfield
   {journal} {\bibinfo  {journal} {Nature Mater.}\ }\textbf {\bibinfo {volume}
  {10}},\ \bibinfo {pages} {419} (\bibinfo {year} {2011})}\BibitemShut
  {NoStop}%
\bibitem [{\citenamefont {Emori}\ \emph {et~al.}(2013)\citenamefont {Emori},
  \citenamefont {Bauer}, \citenamefont {Ahn}, \citenamefont {Martinez},\ and\
  \citenamefont {Beach}}]{Emori2013}%
  \BibitemOpen
  \bibfield  {author} {\bibinfo {author} {\bibfnamefont {S.}~\bibnamefont
  {Emori}}, \bibinfo {author} {\bibfnamefont {U.}~\bibnamefont {Bauer}},
  \bibinfo {author} {\bibfnamefont {S.-M.}\ \bibnamefont {Ahn}}, \bibinfo
  {author} {\bibfnamefont {E.}~\bibnamefont {Martinez}}, \ and\ \bibinfo
  {author} {\bibfnamefont {G.~S.~D.}\ \bibnamefont {Beach}},\ }\href {\doibase
  {10.1038/NMAT3675}} {\bibfield  {journal} {\bibinfo  {journal} {Nature
  Mater.}\ }\textbf {\bibinfo {volume} {12}},\ \bibinfo {pages} {611} (\bibinfo
  {year} {2013})}\BibitemShut {NoStop}%
\bibitem [{\citenamefont {Ryu}\ \emph {et~al.}(2013)\citenamefont {Ryu},
  \citenamefont {Thomas}, \citenamefont {Yang},\ and\ \citenamefont
  {Parkin}}]{Ryu2013}%
  \BibitemOpen
  \bibfield  {author} {\bibinfo {author} {\bibfnamefont {K.-S.}\ \bibnamefont
  {Ryu}}, \bibinfo {author} {\bibfnamefont {L.}~\bibnamefont {Thomas}},
  \bibinfo {author} {\bibfnamefont {S.-H.}\ \bibnamefont {Yang}}, \ and\
  \bibinfo {author} {\bibfnamefont {S.}~\bibnamefont {Parkin}},\ }\href
  {\doibase 10.1038/NNANO.2013.102} {\bibfield  {journal} {\bibinfo  {journal}
  {Nature Nanotech.}\ }\textbf {\bibinfo {volume} {8}},\ \bibinfo {pages} {527}
  (\bibinfo {year} {2013})}\BibitemShut {NoStop}%
\bibitem [{\citenamefont {Je}\ \emph {et~al.}(2013)\citenamefont {Je},
  \citenamefont {Kim}, \citenamefont {Yoo}, \citenamefont {Min}, \citenamefont
  {Lee},\ and\ \citenamefont {Choe}}]{Je2013}%
  \BibitemOpen
  \bibfield  {author} {\bibinfo {author} {\bibfnamefont {S.-G.}\ \bibnamefont
  {Je}}, \bibinfo {author} {\bibfnamefont {D.-H.}\ \bibnamefont {Kim}},
  \bibinfo {author} {\bibfnamefont {S.-C.}\ \bibnamefont {Yoo}}, \bibinfo
  {author} {\bibfnamefont {B.-C.}\ \bibnamefont {Min}}, \bibinfo {author}
  {\bibfnamefont {K.-J.}\ \bibnamefont {Lee}}, \ and\ \bibinfo {author}
  {\bibfnamefont {S.-B.}\ \bibnamefont {Choe}},\ }\href {\doibase
  10.1103/PhysRevB.88.214401} {\bibfield  {journal} {\bibinfo  {journal} {Phys.
  Rev. B}\ }\textbf {\bibinfo {volume} {88}},\ \bibinfo {pages} {214401}
  (\bibinfo {year} {2013})}\BibitemShut {NoStop}%
\bibitem [{\citenamefont {Thiaville}\ \emph {et~al.}(2012)\citenamefont
  {Thiaville}, \citenamefont {Rohart}, \citenamefont {Ju{\'{e}}}, \citenamefont
  {Cros},\ and\ \citenamefont {Fert}}]{Thiaville2012}%
  \BibitemOpen
  \bibfield  {author} {\bibinfo {author} {\bibfnamefont {A.}~\bibnamefont
  {Thiaville}}, \bibinfo {author} {\bibfnamefont {S.}~\bibnamefont {Rohart}},
  \bibinfo {author} {\bibfnamefont {E.}~\bibnamefont {Ju{\'{e}}}}, \bibinfo
  {author} {\bibfnamefont {V.}~\bibnamefont {Cros}}, \ and\ \bibinfo {author}
  {\bibfnamefont {A.}~\bibnamefont {Fert}},\ }\href {\doibase
  10.1209/0295-5075/100/57002} {\bibfield  {journal} {\bibinfo  {journal}
  {EPL}\ }\textbf {\bibinfo {volume} {100}},\ \bibinfo {pages} {57002}
  (\bibinfo {year} {2012})}\BibitemShut {NoStop}%
\bibitem [{\citenamefont {Chen}\ \emph
  {et~al.}(2013{\natexlab{a}})\citenamefont {Chen}, \citenamefont {Ma},
  \citenamefont {N'Diaye}, \citenamefont {Kwon}, \citenamefont {Won},
  \citenamefont {Wu},\ and\ \citenamefont {Schmid}}]{Chen2013}%
  \BibitemOpen
  \bibfield  {author} {\bibinfo {author} {\bibfnamefont {G.}~\bibnamefont
  {Chen}}, \bibinfo {author} {\bibfnamefont {T.}~\bibnamefont {Ma}}, \bibinfo
  {author} {\bibfnamefont {A.~T.}\ \bibnamefont {N'Diaye}}, \bibinfo {author}
  {\bibfnamefont {H.}~\bibnamefont {Kwon}}, \bibinfo {author} {\bibfnamefont
  {C.}~\bibnamefont {Won}}, \bibinfo {author} {\bibfnamefont {Y.}~\bibnamefont
  {Wu}}, \ and\ \bibinfo {author} {\bibfnamefont {A.~K.}\ \bibnamefont
  {Schmid}},\ }\href@noop {} {\bibfield  {journal} {\bibinfo  {journal} {Nature
  Commun.}\ }\textbf {\bibinfo {volume} {4}},\ \bibinfo {pages} {2671}
  (\bibinfo {year} {2013}{\natexlab{a}})}\BibitemShut {NoStop}%
\bibitem [{\citenamefont {Chen}\ \emph
  {et~al.}(2013{\natexlab{b}})\citenamefont {Chen}, \citenamefont {Zhu},
  \citenamefont {Quesada}, \citenamefont {Li}, \citenamefont {N'Diaye},
  \citenamefont {Huo}, \citenamefont {Ma}, \citenamefont {Chen}, \citenamefont
  {Kwon}, \citenamefont {Won}, \citenamefont {Qiu}, \citenamefont {Schmid},\
  and\ \citenamefont {Wu}}]{Chen2013b}%
  \BibitemOpen
  \bibfield  {author} {\bibinfo {author} {\bibfnamefont {G.}~\bibnamefont
  {Chen}}, \bibinfo {author} {\bibfnamefont {J.}~\bibnamefont {Zhu}}, \bibinfo
  {author} {\bibfnamefont {A.}~\bibnamefont {Quesada}}, \bibinfo {author}
  {\bibfnamefont {J.}~\bibnamefont {Li}}, \bibinfo {author} {\bibfnamefont
  {A.~T.}\ \bibnamefont {N'Diaye}}, \bibinfo {author} {\bibfnamefont
  {Y.}~\bibnamefont {Huo}}, \bibinfo {author} {\bibfnamefont {T.~P.}\
  \bibnamefont {Ma}}, \bibinfo {author} {\bibfnamefont {Y.}~\bibnamefont
  {Chen}}, \bibinfo {author} {\bibfnamefont {H.~Y.}\ \bibnamefont {Kwon}},
  \bibinfo {author} {\bibfnamefont {C.}~\bibnamefont {Won}}, \bibinfo {author}
  {\bibfnamefont {Z.~Q.}\ \bibnamefont {Qiu}}, \bibinfo {author} {\bibfnamefont
  {A.~K.}\ \bibnamefont {Schmid}}, \ and\ \bibinfo {author} {\bibfnamefont
  {Y.~Z.}\ \bibnamefont {Wu}},\ }\href {\doibase
  10.1103/PhysRevLett.110.177204} {\bibfield  {journal} {\bibinfo  {journal}
  {Phys. Rev. Lett.}\ }\textbf {\bibinfo {volume} {110}},\ \bibinfo {pages}
  {177204} (\bibinfo {year} {2013}{\natexlab{b}})}\BibitemShut {NoStop}%
\bibitem [{\citenamefont {Manchon}\ \emph {et~al.}(2008)\citenamefont
  {Manchon}, \citenamefont {Pizzini}, \citenamefont {Uhl\'{\i}\u{r}},
  \citenamefont {Vogel}, \citenamefont {Lombard}, \citenamefont {Ducruet},
  \citenamefont {Auffret}, \citenamefont {Rodmacq}, \citenamefont {Dieny},
  \citenamefont {Hochstrasser},\ and\ \citenamefont
  {Panaccione}}]{Manchon2008b}%
  \BibitemOpen
  \bibfield  {author} {\bibinfo {author} {\bibfnamefont {A.}~\bibnamefont
  {Manchon}}, \bibinfo {author} {\bibfnamefont {S.}~\bibnamefont {Pizzini}},
  \bibinfo {author} {\bibfnamefont {V.}~\bibnamefont {Uhl\'{\i}\u{r}}},
  \bibinfo {author} {\bibfnamefont {J.}~\bibnamefont {Vogel}}, \bibinfo
  {author} {\bibfnamefont {L.}~\bibnamefont {Lombard}}, \bibinfo {author}
  {\bibfnamefont {C.}~\bibnamefont {Ducruet}}, \bibinfo {author} {\bibfnamefont
  {S.}~\bibnamefont {Auffret}}, \bibinfo {author} {\bibfnamefont
  {B.}~\bibnamefont {Rodmacq}}, \bibinfo {author} {\bibfnamefont
  {B.}~\bibnamefont {Dieny}}, \bibinfo {author} {\bibfnamefont
  {M.}~\bibnamefont {Hochstrasser}}, \ and\ \bibinfo {author} {\bibfnamefont
  {G.}~\bibnamefont {Panaccione}},\ }\href@noop {} {\bibfield  {journal}
  {\bibinfo  {journal} {J. Magn. Magn. Mater.}\ }\textbf {\bibinfo {volume}
  {320}},\ \bibinfo {pages} {1889} (\bibinfo {year} {2008})}\BibitemShut
  {NoStop}%
\bibitem [{SI-()}]{SI-NucleDM}%
  \BibitemOpen
  \href@noop {} {}\bibinfo {note} {See Supplemental Materials}\BibitemShut {Stop}%
\bibitem [{\citenamefont {Thiaville}(1998)}]{Thiaville1998}%
  \BibitemOpen
  \bibfield  {author} {\bibinfo {author} {\bibfnamefont {A.}~\bibnamefont
  {Thiaville}},\ }\href {\doibase
  http://dx.doi.org/10.1016/S0304-8853(97)01014-7} {\bibfield  {journal}
  {\bibinfo  {journal} {J. Magn. Magn. Mater.}\ }\textbf {\bibinfo {volume}
  {182}},\ \bibinfo {pages} {5} (\bibinfo {year} {1998})}\BibitemShut {NoStop}%
\bibitem [{\citenamefont {Jamet}\ \emph {et~al.}(2001)\citenamefont {Jamet},
  \citenamefont {Wernsdorfer}, \citenamefont {Thirion}, \citenamefont {Mailly},
  \citenamefont {Dupuis}, \citenamefont {M{\'{e}}linon},\ and\ \citenamefont
  {P{\'{e}}rez}}]{Jamet2001}%
  \BibitemOpen
  \bibfield  {author} {\bibinfo {author} {\bibfnamefont {M.}~\bibnamefont
  {Jamet}}, \bibinfo {author} {\bibfnamefont {W.}~\bibnamefont {Wernsdorfer}},
  \bibinfo {author} {\bibfnamefont {C.}~\bibnamefont {Thirion}}, \bibinfo
  {author} {\bibfnamefont {D.}~\bibnamefont {Mailly}}, \bibinfo {author}
  {\bibfnamefont {V.}~\bibnamefont {Dupuis}}, \bibinfo {author} {\bibfnamefont
  {P.}~\bibnamefont {M{\'{e}}linon}}, \ and\ \bibinfo {author} {\bibfnamefont
  {A.}~\bibnamefont {P{\'{e}}rez}},\ }\href@noop {} {\bibfield  {journal}
  {\bibinfo  {journal} {Phys.\ Rev.\ Lett.}\ }\textbf {\bibinfo {volume}
  {86}},\ \bibinfo {pages} {4676} (\bibinfo {year} {2001})}\BibitemShut
  {NoStop}%
\bibitem [{\citenamefont {Vouille}\ \emph {et~al.}(2004)\citenamefont
  {Vouille}, \citenamefont {Thiaville},\ and\ \citenamefont
  {Miltat}}]{Vouille2004}%
  \BibitemOpen
  \bibfield  {author} {\bibinfo {author} {\bibfnamefont {C.}~\bibnamefont
  {Vouille}}, \bibinfo {author} {\bibfnamefont {A.}~\bibnamefont {Thiaville}},
  \ and\ \bibinfo {author} {\bibfnamefont {J.}~\bibnamefont {Miltat}},\
  }\href@noop {} {\bibfield  {journal} {\bibinfo  {journal} {J. Magn. Magn.
  Mater.}\ }\textbf {\bibinfo {volume} {272}},\ \bibinfo {pages} {E1237}
  (\bibinfo {year} {2004})}\BibitemShut {NoStop}%
\bibitem [{\citenamefont {Buda}\ \emph {et~al.}(2002)\citenamefont {Buda},
  \citenamefont {Prejbeanu}, \citenamefont {Ebels},\ and\ \citenamefont
  {Ounadjela}}]{Buda2002}%
  \BibitemOpen
  \bibfield  {author} {\bibinfo {author} {\bibfnamefont {L.~D.}\ \bibnamefont
  {Buda}}, \bibinfo {author} {\bibfnamefont {I.~L.}\ \bibnamefont {Prejbeanu}},
  \bibinfo {author} {\bibfnamefont {U.}~\bibnamefont {Ebels}}, \ and\ \bibinfo
  {author} {\bibfnamefont {K.}~\bibnamefont {Ounadjela}},\ }\href {\doibase
  {10.1016/S0927-0256(02)00184-2}} {\bibfield  {journal} {\bibinfo  {journal}
  {Comput. Mater. Sci.}\ }\textbf {\bibinfo {volume} {24}},\ \bibinfo {pages}
  {181} (\bibinfo {year} {2002})}\BibitemShut {NoStop}%
\bibitem [{\citenamefont {Boulle}\ \emph {et~al.}(2013)\citenamefont {Boulle},
  \citenamefont {Rohart}, \citenamefont {Buda-Prejbeanu}, \citenamefont
  {Ju\'e}, \citenamefont {Miron}, \citenamefont {Pizzini}, \citenamefont
  {Vogel}, \citenamefont {Gaudin},\ and\ \citenamefont
  {Thiaville}}]{Boulle2013}%
  \BibitemOpen
  \bibfield  {author} {\bibinfo {author} {\bibfnamefont {O.}~\bibnamefont
  {Boulle}}, \bibinfo {author} {\bibfnamefont {S.}~\bibnamefont {Rohart}},
  \bibinfo {author} {\bibfnamefont {L.~D.}\ \bibnamefont {Buda-Prejbeanu}},
  \bibinfo {author} {\bibfnamefont {E.}~\bibnamefont {Ju\'e}}, \bibinfo
  {author} {\bibfnamefont {I.~M.}\ \bibnamefont {Miron}}, \bibinfo {author}
  {\bibfnamefont {S.}~\bibnamefont {Pizzini}}, \bibinfo {author} {\bibfnamefont
  {J.}~\bibnamefont {Vogel}}, \bibinfo {author} {\bibfnamefont
  {G.}~\bibnamefont {Gaudin}}, \ and\ \bibinfo {author} {\bibfnamefont
  {A.}~\bibnamefont {Thiaville}},\ }\href {\doibase
  10.1103/PhysRevLett.111.217203} {\bibfield  {journal} {\bibinfo  {journal}
  {Phys. Rev. Lett.}\ }\textbf {\bibinfo {volume} {111}},\ \bibinfo {pages}
  {217203} (\bibinfo {year} {2013})}\BibitemShut {NoStop}%
\bibitem [{\citenamefont {Metaxas}\ \emph {et~al.}(2007)\citenamefont
  {Metaxas}, \citenamefont {Jamet}, \citenamefont {Mougin}, \citenamefont
  {Cormier}, \citenamefont {Ferr\'e}, \citenamefont {Baltz}, \citenamefont
  {Rodmacq}, \citenamefont {Dieny},\ and\ \citenamefont
  {Stamps}}]{Metaxas2007}%
  \BibitemOpen
  \bibfield  {author} {\bibinfo {author} {\bibfnamefont {P.~J.}\ \bibnamefont
  {Metaxas}}, \bibinfo {author} {\bibfnamefont {J.~P.}\ \bibnamefont {Jamet}},
  \bibinfo {author} {\bibfnamefont {A.}~\bibnamefont {Mougin}}, \bibinfo
  {author} {\bibfnamefont {M.}~\bibnamefont {Cormier}}, \bibinfo {author}
  {\bibfnamefont {J.}~\bibnamefont {Ferr\'e}}, \bibinfo {author} {\bibfnamefont
  {V.}~\bibnamefont {Baltz}}, \bibinfo {author} {\bibfnamefont
  {B.}~\bibnamefont {Rodmacq}}, \bibinfo {author} {\bibfnamefont
  {B.}~\bibnamefont {Dieny}}, \ and\ \bibinfo {author} {\bibfnamefont {R.~L.}\
  \bibnamefont {Stamps}},\ }\href {\doibase 10.1103/PhysRevLett.99.217208}
  {\bibfield  {journal} {\bibinfo  {journal} {Phys. Rev. Lett.}\ }\textbf
  {\bibinfo {volume} {99}},\ \bibinfo {pages} {217208} (\bibinfo {year}
  {2007})}\BibitemShut {NoStop}%
\bibitem [{\citenamefont {Aharoni}(1962)}]{Aharoni1962}%
  \BibitemOpen
  \bibfield  {author} {\bibinfo {author} {\bibfnamefont {A.}~\bibnamefont
  {Aharoni}},\ }\href@noop {} {\bibfield  {journal} {\bibinfo  {journal} {Rev.\
  Mod.\ Phys.}\ }\textbf {\bibinfo {volume} {34}},\ \bibinfo {pages} {227}
  (\bibinfo {year} {1962})}\BibitemShut {NoStop}%
\bibitem [{\citenamefont {Rohart}\ and\ \citenamefont
  {Thiaville}(2013)}]{Rohart2013}%
  \BibitemOpen
  \bibfield  {author} {\bibinfo {author} {\bibfnamefont {S.}~\bibnamefont
  {Rohart}}\ and\ \bibinfo {author} {\bibfnamefont {A.}~\bibnamefont
  {Thiaville}},\ }\href {\doibase 10.1103/PhysRevB.88.184422} {\bibfield
  {journal} {\bibinfo  {journal} {Phys. Rev. B}\ }\textbf {\bibinfo {volume}
  {88}},\ \bibinfo {pages} {184422} (\bibinfo {year} {2013})}\BibitemShut
  {NoStop}%
\bibitem [{\citenamefont {Stoner}\ and\ \citenamefont
  {Wohlfarth}(1948)}]{Stoner1948}%
  \BibitemOpen
  \bibfield  {author} {\bibinfo {author} {\bibfnamefont {E.~C.}\ \bibnamefont
  {Stoner}}\ and\ \bibinfo {author} {\bibfnamefont {E.~P.}\ \bibnamefont
  {Wohlfarth}},\ }\href@noop {} {\bibfield  {journal} {\bibinfo  {journal}
  {Phil.\ Trans.\ Roy.\ Soc. London}\ }\textbf {\bibinfo {volume} {A240}},\
  \bibinfo {pages} {599} (\bibinfo {year} {1948})},\ \bibinfo {note} {reprinted
  in IEEE Trans.\ Magn., vol. 27, pp. 3475-3518 (1991)}\BibitemShut {NoStop}%
\bibitem [{\citenamefont {Barbara}\ and\ \citenamefont
  {Uehara}(1976)}]{Barbara1976}%
  \BibitemOpen
  \bibfield  {author} {\bibinfo {author} {\bibfnamefont {B.}~\bibnamefont
  {Barbara}}\ and\ \bibinfo {author} {\bibfnamefont {M.}~\bibnamefont
  {Uehara}},\ }\href@noop {} {\bibfield  {journal} {\bibinfo  {journal} {IEEE
  Trans.\ Magn.}\ }\textbf {\bibinfo {volume} {83}},\ \bibinfo {pages} {997}
  (\bibinfo {year} {1976})}\BibitemShut {NoStop}%
\bibitem [{\citenamefont {Barbara}(1994)}]{Barbara1994}%
  \BibitemOpen
  \bibfield  {author} {\bibinfo {author} {\bibfnamefont {B.}~\bibnamefont
  {Barbara}},\ }\href@noop {} {\bibfield  {journal} {\bibinfo  {journal} {J.
  Magn. Magn. Mater.}\ }\textbf {\bibinfo {volume} {129}},\ \bibinfo {pages}
  {79} (\bibinfo {year} {1994})}\BibitemShut {NoStop}%
\bibitem [{\citenamefont {Moritz}\ \emph {et~al.}(2005)\citenamefont {Moritz},
  \citenamefont {Dieny}, \citenamefont {Nozi{\`{e}}res}, \citenamefont
  {Pennec}, \citenamefont {Camarero},\ and\ \citenamefont
  {Pizzini}}]{Moritz2005}%
  \BibitemOpen
  \bibfield  {author} {\bibinfo {author} {\bibfnamefont {J.}~\bibnamefont
  {Moritz}}, \bibinfo {author} {\bibfnamefont {B.}~\bibnamefont {Dieny}},
  \bibinfo {author} {\bibfnamefont {J.-P.}~\bibnamefont {Nozi{\`{e}}res}},
  \bibinfo {author} {\bibfnamefont {Y.}~\bibnamefont {Pennec}}, \bibinfo
  {author} {\bibfnamefont {J.}~\bibnamefont {Camarero}}, \ and\ \bibinfo
  {author} {\bibfnamefont {S.}~\bibnamefont {Pizzini}},\ }\href {\doibase
  {10.1103/PhysRevB.71.100402}} {\bibfield  {journal} {\bibinfo  {journal}
  {Phys. Rev. B}\ }\textbf {\bibinfo {volume} {71}},\ \bibinfo {pages}
  {{100402}} (\bibinfo {year} {2005})}\BibitemShut {NoStop}%
\bibitem [{\citenamefont {Hubert}(1974)}]{Hubert1974}%
  \BibitemOpen
  \bibfield  {author} {\bibinfo {author} {\bibfnamefont {A.}~\bibnamefont
  {Hubert}},\ }\href@noop {} {\emph {\bibinfo {title} {Theorie der
  Dom{\"{a}}nenw{\"{a}}nde in geordneten Medien (in german)}}}\ (\bibinfo
  {publisher} {Springer Verlag},\ \bibinfo {year} {1974})\BibitemShut {NoStop}%
\bibitem [{\citenamefont {Desjonqu{\`{e}}res}\ and\ \citenamefont
  {Spanjaard}(1996)}]{Desjonqueres1996}%
  \BibitemOpen
  \bibfield  {author} {\bibinfo {author} {\bibfnamefont {M.~C.}\ \bibnamefont
  {Desjonqu{\`{e}}res}}\ and\ \bibinfo {author} {\bibfnamefont
  {D.}~\bibnamefont {Spanjaard}},\ }\href@noop {} {\emph {\bibinfo {title}
  {Concepts in Surface Physics}}},\ Springer Series in Surface Sciences, vol.
  30\ (\bibinfo  {publisher} {Springer, Berlin},\ \bibinfo {year}
  {1996})\BibitemShut {NoStop}%
\bibitem [{not({\natexlab{b}})}]{noteHubert}%
  \BibitemOpen
  \href@noop {} {} \ \bibinfo {note} {the calculation without
  DMI can be found in \cite{Hubert1974}.}\BibitemShut {Stop}%
\bibitem [{\citenamefont {Sampaio}\ \emph {et~al.}(2013)\citenamefont
  {Sampaio}, \citenamefont {Cros}, \citenamefont {Rohart}, \citenamefont
  {Thiaville},\ and\ \citenamefont {Fert}}]{Sampaio2013}%
  \BibitemOpen
  \bibfield  {author} {\bibinfo {author} {\bibfnamefont {J.}~\bibnamefont
  {Sampaio}}, \bibinfo {author} {\bibfnamefont {V.}~\bibnamefont {Cros}},
  \bibinfo {author} {\bibfnamefont {S.}~\bibnamefont {Rohart}}, \bibinfo
  {author} {\bibfnamefont {A.}~\bibnamefont {Thiaville}}, \ and\ \bibinfo
  {author} {\bibfnamefont {A.}~\bibnamefont {Fert}},\ }\href {\doibase
  {10.1038/NNANO.2013.210}} {\bibfield  {journal} {\bibinfo  {journal} {Nature
  Nanotech.}\ }\textbf {\bibinfo {volume} {8}},\ \bibinfo {pages} {839}
  (\bibinfo {year} {2013})}\BibitemShut {NoStop}%
\bibitem [{\citenamefont {Lee}\ \emph {et~al.}(2014)\citenamefont {Lee},
  \citenamefont {Liu}, \citenamefont {Pai}, \citenamefont {Li}, \citenamefont
  {Tseng}, \citenamefont {Gowtham}, \citenamefont {Park}, \citenamefont
  {Ralph},\ and\ \citenamefont {Buhrman}}]{LeePRB2014}%
  \BibitemOpen
  \bibfield  {author} {\bibinfo {author} {\bibfnamefont {O.~J.}\ \bibnamefont
  {Lee}}, \bibinfo {author} {\bibfnamefont {L.~Q.}\ \bibnamefont {Liu}},
  \bibinfo {author} {\bibfnamefont {C.~F.}\ \bibnamefont {Pai}}, \bibinfo
  {author} {\bibfnamefont {Y.}~\bibnamefont {Li}}, \bibinfo {author}
  {\bibfnamefont {H.~W.}\ \bibnamefont {Tseng}}, \bibinfo {author}
  {\bibfnamefont {P.~G.}\ \bibnamefont {Gowtham}}, \bibinfo {author}
  {\bibfnamefont {J.~P.}\ \bibnamefont {Park}}, \bibinfo {author}
  {\bibfnamefont {D.~C.}\ \bibnamefont {Ralph}}, \ and\ \bibinfo {author}
  {\bibfnamefont {R.~A.}\ \bibnamefont {Buhrman}},\ }\href@noop {} {\bibfield
  {journal} {\bibinfo  {journal} {Phys.\ Rev.\ B}\ }\textbf {\bibinfo {volume}
  {89}},\ \bibinfo {pages} {024418} (\bibinfo {year} {2014})}\BibitemShut
  {NoStop}%
\end{thebibliography}

\begin{thebibliography}{10}%
\makeatletter
\providecommand \@ifxundefined [1]{%
 \@ifx{#1\undefined}
}%
\providecommand \@ifnum [1]{%
 \ifnum #1\expandafter \@firstoftwo
 \else \expandafter \@secondoftwo
 \fi
}%
\providecommand \@ifx [1]{%
 \ifx #1\expandafter \@firstoftwo
 \else \expandafter \@secondoftwo
 \fi
}%
\providecommand \natexlab [1]{#1}%
\providecommand \enquote  [1]{``#1''}%
\providecommand \bibnamefont  [1]{#1}%
\providecommand \bibfnamefont [1]{#1}%
\providecommand \citenamefont [1]{#1}%
\providecommand \href@noop [0]{\@secondoftwo}%
\providecommand \href [0]{\begingroup \@sanitize@url \@href}%
\providecommand \@href[1]{\@@startlink{#1}\@@href}%
\providecommand \@@href[1]{\endgroup#1\@@endlink}%
\providecommand \@sanitize@url [0]{\catcode `\\12\catcode `\$12\catcode
  `\&12\catcode `\#12\catcode `\^12\catcode `\_12\catcode `\%12\relax}%
\providecommand \@@startlink[1]{}%
\providecommand \@@endlink[0]{}%
\providecommand \url  [0]{\begingroup\@sanitize@url \@url }%
\providecommand \@url [1]{\endgroup\@href {#1}{\urlprefix }}%
\providecommand \urlprefix  [0]{URL }%
\providecommand \Eprint [0]{\href }%
\providecommand \doibase [0]{http://dx.doi.org/}%
\providecommand \selectlanguage [0]{\@gobble}%
\providecommand \bibinfo  [0]{\@secondoftwo}%
\providecommand \bibfield  [0]{\@secondoftwo}%
\providecommand \translation [1]{[#1]}%
\providecommand \BibitemOpen [0]{}%
\providecommand \bibitemStop [0]{}%
\providecommand \bibitemNoStop [0]{.\EOS\space}%
\providecommand \EOS [0]{\spacefactor3000\relax}%
\providecommand \BibitemShut  [1]{\csname bibitem#1\endcsname}%
\let\auto@bib@innerbib\@empty
\bibitem [{\citenamefont {Stoner}\ and\ \citenamefont
  {Wohlfarth}(1948)}]{Stoner1948SI}%
  \BibitemOpen
  \bibfield  {author} {\bibinfo {author} {\bibfnamefont {E.~C.}\ \bibnamefont
  {Stoner}}\ and\ \bibinfo {author} {\bibfnamefont {E.~P.}\ \bibnamefont
  {Wohlfarth}},\ }\href@noop {} {\bibfield  {journal} {\bibinfo  {journal}
  {Phil.\ Trans.\ Roy.\ Soc. London}\ }\textbf {\bibinfo {volume} {A240}},\
  \bibinfo {pages} {599} (\bibinfo {year} {1948})},\ \bibinfo {note} {reprinted
  in IEEE Trans.\ Magn., vol. 27, pp. 3475-3518 (1991)}\BibitemShut {NoStop}%
\bibitem [{\citenamefont {Thiaville}(1998)}]{Thiaville1998SI}%
  \BibitemOpen
  \bibfield  {author} {\bibinfo {author} {\bibfnamefont {A.}~\bibnamefont
  {Thiaville}},\ }\href {\doibase
  http://dx.doi.org/10.1016/S0304-8853(97)01014-7} {\bibfield  {journal}
  {\bibinfo  {journal} {J. Magn. Magn. Mater.}\ }\textbf {\bibinfo {volume}
  {182}},\ \bibinfo {pages} {5} (\bibinfo {year} {1998})}\BibitemShut {NoStop}%
\bibitem [{\citenamefont {Rohart}\ and\ \citenamefont
  {Thiaville}(2013)}]{Rohart2013a}%
  \BibitemOpen
  \bibfield  {author} {\bibinfo {author} {\bibfnamefont {S.}~\bibnamefont
  {Rohart}}\ and\ \bibinfo {author} {\bibfnamefont {A.}~\bibnamefont
  {Thiaville}},\ }\href {\doibase 10.1103/PhysRevB.88.184422} {\bibfield
  {journal} {\bibinfo  {journal} {Phys. Rev. B}\ }\textbf {\bibinfo {volume}
  {88}},\ \bibinfo {pages} {184422} (\bibinfo {year} {2013})}\BibitemShut
  {NoStop}%
\bibitem [{\citenamefont {Jamet}\ \emph {et~al.}(2001)\citenamefont {Jamet},
  \citenamefont {Wernsdorfer}, \citenamefont {Thirion}, \citenamefont {Mailly},
  \citenamefont {Dupuis}, \citenamefont {M{\'{e}}linon},\ and\ \citenamefont
  {P{\'{e}}rez}}]{Jamet2001SI}%
  \BibitemOpen
  \bibfield  {author} {\bibinfo {author} {\bibfnamefont {M.}~\bibnamefont
  {Jamet}}, \bibinfo {author} {\bibfnamefont {W.}~\bibnamefont {Wernsdorfer}},
  \bibinfo {author} {\bibfnamefont {C.}~\bibnamefont {Thirion}}, \bibinfo
  {author} {\bibfnamefont {D.}~\bibnamefont {Mailly}}, \bibinfo {author}
  {\bibfnamefont {V.}~\bibnamefont {Dupuis}}, \bibinfo {author} {\bibfnamefont
  {P.}~\bibnamefont {M{\'{e}}linon}}, \ and\ \bibinfo {author} {\bibfnamefont
  {A.}~\bibnamefont {P{\'{e}}rez}},\ }\href@noop {} {\bibfield  {journal}
  {\bibinfo  {journal} {Phys.\ Rev.\ Lett.}\ }\textbf {\bibinfo {volume}
  {86}},\ \bibinfo {pages} {4676} (\bibinfo {year} {2001})}\BibitemShut
  {NoStop}%
\bibitem [{\citenamefont {Vouille}\ \emph {et~al.}(2004)\citenamefont
  {Vouille}, \citenamefont {Thiaville},\ and\ \citenamefont
  {Miltat}}]{Vouille2004SI}%
  \BibitemOpen
  \bibfield  {author} {\bibinfo {author} {\bibfnamefont {C.}~\bibnamefont
  {Vouille}}, \bibinfo {author} {\bibfnamefont {A.}~\bibnamefont {Thiaville}},
  \ and\ \bibinfo {author} {\bibfnamefont {J.}~\bibnamefont {Miltat}},\
  }\href@noop {} {\bibfield  {journal} {\bibinfo  {journal} {J. Magn. Magn.
  Mater.}\ }\textbf {\bibinfo {volume} {272}},\ \bibinfo {pages} {E1237}
  (\bibinfo {year} {2004})}\BibitemShut {NoStop}%
\bibitem [{\citenamefont {Thiaville}\ \emph {et~al.}(2012)\citenamefont
  {Thiaville}, \citenamefont {Rohart}, \citenamefont {Ju{\'{e}}}, \citenamefont
  {Cros},\ and\ \citenamefont {Fert}}]{Thiaville2012a}%
  \BibitemOpen
  \bibfield  {author} {\bibinfo {author} {\bibfnamefont {A.}~\bibnamefont
  {Thiaville}}, \bibinfo {author} {\bibfnamefont {S.}~\bibnamefont {Rohart}},
  \bibinfo {author} {\bibfnamefont {E.}~\bibnamefont {Ju{\'{e}}}}, \bibinfo
  {author} {\bibfnamefont {V.}~\bibnamefont {Cros}}, \ and\ \bibinfo {author}
  {\bibfnamefont {A.}~\bibnamefont {Fert}},\ }\href {\doibase
  10.1209/0295-5075/100/57002} {\bibfield  {journal} {\bibinfo  {journal}
  {EPL}\ }\textbf {\bibinfo {volume} {100}},\ \bibinfo {pages} {57002}
  (\bibinfo {year} {2012})}\BibitemShut {NoStop}%
\bibitem [{\citenamefont {Buda}\ \emph {et~al.}(2002)\citenamefont {Buda},
  \citenamefont {Prejbeanu}, \citenamefont {Ebels},\ and\ \citenamefont
  {Ounadjela}}]{Buda2002SI}%
  \BibitemOpen
  \bibfield  {author} {\bibinfo {author} {\bibfnamefont {L.~D.}\ \bibnamefont
  {Buda}}, \bibinfo {author} {\bibfnamefont {I.~L.}\ \bibnamefont {Prejbeanu}},
  \bibinfo {author} {\bibfnamefont {U.}~\bibnamefont {Ebels}}, \ and\ \bibinfo
  {author} {\bibfnamefont {K.}~\bibnamefont {Ounadjela}},\ }\href {\doibase
  {10.1016/S0927-0256(02)00184-2}} {\bibfield  {journal} {\bibinfo  {journal}
  {Comput. Mater. Sci.}\ }\textbf {\bibinfo {volume} {24}},\ \bibinfo {pages}
  {181} (\bibinfo {year} {2002})}\BibitemShut {NoStop}%
\bibitem [{\citenamefont {Boulle}\ \emph {et~al.}(2013)\citenamefont {Boulle},
  \citenamefont {Rohart}, \citenamefont {Buda-Prejbeanu}, \citenamefont
  {Ju\'e}, \citenamefont {Miron}, \citenamefont {Pizzini}, \citenamefont
  {Vogel}, \citenamefont {Gaudin},\ and\ \citenamefont
  {Thiaville}}]{Boulle2013SI}%
  \BibitemOpen
  \bibfield  {author} {\bibinfo {author} {\bibfnamefont {O.}~\bibnamefont
  {Boulle}}, \bibinfo {author} {\bibfnamefont {S.}~\bibnamefont {Rohart}},
  \bibinfo {author} {\bibfnamefont {L.~D.}\ \bibnamefont {Buda-Prejbeanu}},
  \bibinfo {author} {\bibfnamefont {E.}~\bibnamefont {Ju\'e}}, \bibinfo
  {author} {\bibfnamefont {I.~M.}\ \bibnamefont {Miron}}, \bibinfo {author}
  {\bibfnamefont {S.}~\bibnamefont {Pizzini}}, \bibinfo {author} {\bibfnamefont
  {J.}~\bibnamefont {Vogel}}, \bibinfo {author} {\bibfnamefont
  {G.}~\bibnamefont {Gaudin}}, \ and\ \bibinfo {author} {\bibfnamefont
  {A.}~\bibnamefont {Thiaville}},\ }\href {\doibase
  10.1103/PhysRevLett.111.217203} {\bibfield  {journal} {\bibinfo  {journal}
  {Phys. Rev. Lett.}\ }\textbf {\bibinfo {volume} {111}},\ \bibinfo {pages}
  {217203} (\bibinfo {year} {2013})}\BibitemShut {NoStop}%
\bibitem [{\citenamefont {Metaxas}\ \emph {et~al.}(2007)\citenamefont
  {Metaxas}, \citenamefont {Jamet}, \citenamefont {Mougin}, \citenamefont
  {Cormier}, \citenamefont {Ferr\'e}, \citenamefont {Baltz}, \citenamefont
  {Rodmacq}, \citenamefont {Dieny},\ and\ \citenamefont
  {Stamps}}]{Metaxas2007SI}%
  \BibitemOpen
  \bibfield  {author} {\bibinfo {author} {\bibfnamefont {P.~J.}\ \bibnamefont
  {Metaxas}}, \bibinfo {author} {\bibfnamefont {J.~P.}\ \bibnamefont {Jamet}},
  \bibinfo {author} {\bibfnamefont {A.}~\bibnamefont {Mougin}}, \bibinfo
  {author} {\bibfnamefont {M.}~\bibnamefont {Cormier}}, \bibinfo {author}
  {\bibfnamefont {J.}~\bibnamefont {Ferr\'e}}, \bibinfo {author} {\bibfnamefont
  {V.}~\bibnamefont {Baltz}}, \bibinfo {author} {\bibfnamefont
  {B.}~\bibnamefont {Rodmacq}}, \bibinfo {author} {\bibfnamefont
  {B.}~\bibnamefont {Dieny}}, \ and\ \bibinfo {author} {\bibfnamefont {R.~L.}\
  \bibnamefont {Stamps}},\ }\href {\doibase 10.1103/PhysRevLett.99.217208}
  {\bibfield  {journal} {\bibinfo  {journal} {Phys. Rev. Lett.}\ }\textbf
  {\bibinfo {volume} {99}},\ \bibinfo {pages} {217208} (\bibinfo {year}
  {2007})}\BibitemShut {NoStop}%
\bibitem [{\citenamefont {Moritz}\ \emph {et~al.}(2005)\citenamefont {Moritz},
  \citenamefont {Dieny}, \citenamefont {Nozi{\`{e}}res}, \citenamefont
  {Pennec}, \citenamefont {Camarero},\ and\ \citenamefont
  {Pizzini}}]{Moritz2005a}%
  \BibitemOpen
  \bibfield  {author} {\bibinfo {author} {\bibfnamefont {J.}~\bibnamefont
  {Moritz}}, \bibinfo {author} {\bibfnamefont {B.}~\bibnamefont {Dieny}},
  \bibinfo {author} {\bibfnamefont {J.}~\bibnamefont {Nozi{\`{e}}res}},
  \bibinfo {author} {\bibfnamefont {Y.}~\bibnamefont {Pennec}}, \bibinfo
  {author} {\bibfnamefont {J.}~\bibnamefont {Camarero}}, \ and\ \bibinfo
  {author} {\bibfnamefont {S.}~\bibnamefont {Pizzini}},\ }\href {\doibase
  {10.1103/PhysRevB.71.100402}} {\bibfield  {journal} {\bibinfo  {journal}
  {Phys. Rev. B}\ }\textbf {\bibinfo {volume} {71}},\ \bibinfo {pages}
  {{100402}} (\bibinfo {year} {2005})}\BibitemShut {NoStop}%
\end{thebibliography}
\end{document}